\documentclass{pasj00}
\begin{document}
\SetRunningHead{C. Hikage et al.}{Fourier Phase Analysis of SDSS
Galaxies} \Received{2005/06/09}%{yyyy/mm/dd}
\Accepted{2005/08/11}%{yyyy/mm/dd}

\title{Fourier Phase Analysis of SDSS Galaxies}

%%% begin:list of authors
\author{%
Chiaki \textsc{Hikage} and Takahiko \textsc{Matsubara}}
\affil{Department of Physics and Astrophysics, 
Nagoya University, Chikusa, Nagoya 464-8602}
\email{hikage@a.phys.nagoya-u.ac.jp, 
taka@a.phys.nagoya-u.ac.jp}
\author{Yasushi \textsc{Suto}}
\affil{Department of Physics, School of Science, 
The University of Tokyo, Tokyo 113-0033}
\email{suto@phys.s.u-tokyo.ac.jp}
\author{Changbom \textsc{Park}}
\affil{Korea Institute for Advanced Study, Dongdaemun-gu, Seoul
207-43, Korea}
\email{cbp@kias.re.kr}
\author{Alexander S. \textsc{Szalay}}
\affil{Department of Physics and Astronomy, The Johns Hopkins University,
Baltimore, MD 21218, USA}
\email{szalay@jhu.edu}
\and \author{Jon \textsc{Brinkmann}}
\affil{Apache Point Observatory, P.O.Box 59, Sunspot, NM
88349-0059, USA}
\email{jb@apo.nmsu.edu}
\maketitle

\begin{abstract}
We present a first analysis of the clustering of SDSS galaxies using
the distribution function of the sum of Fourier phases. This
statistical method was recently proposed by one of the authors as a
new probe of the phase correlations of cosmological density
fields. Since the Fourier phases are statistically independent of the
Fourier amplitudes, the phase statistic plays a complementary role to
the conventional two-point statistics of galaxy clustering.  In
particular, we focus on the distribution functions of the phase sum
over three closed wavevectors as a function of the triangle
configuration.  We find that the observed distribution functions of
the phase sum are in good agreement with the lowest-order
approximation from perturbation theory. For a direct comparison with
observations, we construct mock catalogs from $N$-body simulations
taking account of the survey geometry, the redshift distortion, and
the discreteness due to the limited number of data. Indeed the
observed phase correlations for the galaxies in the range of the
absolute magnitude, $-22<M_r<-18$, agree well with those for
$\Lambda$-dominated spatially flat cold dark matter predictions with
$\sigma_8=0.9$ evolved from the Gaussian initial condition.  This
agreement implies that the galaxy biasing is approximately linear in
redshift space. Instead, assuming that the galaxy biasing is described
by a quadratic deterministic function at $k<0.03$[$2\pi/(h^{-1}{\rm
Mpc})$], we can constrain the ratio of the quadratic biasing
parameter, $b_2$, to the linear biasing parameter, $b_1$, from the
difference of phase correlations between observations and mock
predictions. We find that the resulting $b_2/b_1$ is well fitted by
$b_2/b_1=0.54(\pm 0.06)-0.62(\pm 0.08)\sigma_8$ and is almost
insensitive to the cosmology and luminosity in those ranges.  Indeed,
$b_2/b_1$ is nearly zero when $\sigma_8=0.9$.
\end{abstract}

\newpage
\section{Introduction}
\label{sec:intro}

The spatial distribution of galaxies, {\it Large Scale Structure} (LSS),
provides fundamental knowledge about the formation and evolution of
the density structure in our universe.  A recent measurement of the CMB
(Cosmological Microwave Background) anisotropy by WMAP (Wilkinson
Microwave Anisotropy Probe) puts a stringent constraint on the primordial
density fluctuations, and favors the $\Lambda$-dominated spatially flat
Cold Dark Matter (LCDM) model with a Gaussian initial condition
\citep{Spergel2003,Komatsu2003}.  One of the goals in the analysis of
galaxy distribution is to test cosmological models in an independent and
complementary manner to the CMB analysis. Another goal is to understand
galaxy clustering, with particular emphasis on galaxy biasing,
which is a statistical relation of the clustering between galaxies and
the underlying dark matter. Recent wide-field galaxy surveys, such as
SDSS (Sloan Digital Sky Survey) and 2dFGRS (Two Degree Field Galaxy
Redshift Survey), indeed enable a detailed study of galaxy clustering
with unprecedented accuracy.

More conventional statistics for analyzing the galaxy distribution is
the two-point correlation function, or the power spectrum in Fourier
space. Two-point statistics have been extensively studied for more
than 35 years since the pioneer work by \citet{TK1969}, and have been
applied to various redshift surveys, including SDSS (e.g., 
\cite{Tegmark2004a,Zehavi2005} for SDSS galaxies; \cite{Yahata2005}
for SDSS quasars).  They, however, cannot fully describe the
statistical nature of the galaxy distribution, because LSS is highly
non-Gaussian due to the nonlinear gravitational evolution and
nonlinearity in the galaxy biasing. To characterize the non-Gaussian
properties of the galaxy distribution requires higher order statistics
beyond two-point statistics.  Motivated by this requirement, various
statistics have been introduced as complementary tools to two-point
statistics for the analysis of galaxy clustering: higher order
correlation functions (e.g., Peebles 1980), genus statistics
\citep{GMD1986,Hoyle2002,Hikage2002,Park2005}, Minkowski functionals
\citep{MBW1994,Hikage2003}, minimum spanning trees \citep{BBS1985}, void
statistics \citep{White1979}, and so on. While these statistics
capture different parts of non-Gaussian features, they are not
completely independent of the two-point statistics.

The Fourier transform, $\delta_{\scriptsize\mathbf k}$, of a density
fluctuation field is separately written in terms of the amplitude and
the phase, $\theta_{\scriptsize\mathbf k}$, as
%%%%%%%%%%%%%%%%%%%%%%%%%%%%%%%%%%%%%%%%%%%%%%%%%%%%%%%
\begin{equation}
\delta_{\scriptsize\mathbf k}=|\delta_{\scriptsize\mathbf
k}|\exp(i\theta_{\scriptsize\mathbf k}).
\end{equation}
%%%%%%%%%%%%%%%%%%%%%%%%%%%%%%%%%%%%%%%%%%%%%%%%%%%%%%%
The Gaussian fields have a uniform distribution of the Fourier phases
over $0\le\theta_{\scriptsize\mathbf k}\le 2\pi$.  Therefore,
characterizing the correlation of phases is expected to be a direct
means to explore the non-Gaussian features. Furthermore, Fourier
phases are statistically independent information of the power
spectrum, which is defined by $|\delta_{\scriptsize\mathbf
k}|^2$. Nevertheless, finding useful statistics of the Fourier phases
is not easy, mainly because of the cyclic property of the phase.  For
example, the one-point phase distribution turns out to be essentially
uniform, even in a strongly non-Gaussian field \citep{Suginohara1991}. 
For this reason, previous studies of the Fourier phase have been
mainly devoted to the evolution of phase shifts in individual modes
\citep{Ryden1991, Soda1992, Jain1998}, or the phase differences
between the Fourier modes \citep{SMS1991, CC2000, Chiang2001, CCN2002,
WC2003}.

\citet{Matsu2003} recently proposed, as a new measure of phase
correlations, the distribution function of the ``phase sum'',
$\theta_{{\scriptsize\mathbf k}_1}+\theta_{{\scriptsize\mathbf
k}_2}+\cdots + \theta_{{\scriptsize\mathbf k}_N}$, where the
corresponding wavevectors satisfy ${\mathbf k}_1+{\mathbf k}_2+ \cdots
+ {\mathbf k}_N={\mathbf 0}$.  Although a connection between the
higher order statistics and the phase correlations was suggested
earlier \citep{Bertschinger1992, WC2003}, he discovered an important
analytic relation between the distribution function of the phase sum
and the polyspectra using perturbation theory.

A subsequent numerical analysis by \citet{Hikage2004} explained the
behavior of the phase-sum distribution with respect to the density
structure in real space. When a prominent density peak exists in a
given sampling volume, the Fourier phases of various modes are
synchronized to have nearly zero values at the position of peak, and
thus the phase sum distributes around zero. On the other hand, when
several peaks with comparable heights exist, the synchronization of
phases is diluted, and thus the phase-sum distribution becomes almost
uniform.  The phase-sum distribution is sensitive to the non-Gaussian
feature of the density field, especially the relative strength of the
most high-density peak to other density peaks.  The nonlinear
gravitational evolution and the nonlinearity in the galaxy biasing
statistically changes the relative strength of density peaks in galaxy
distribution, and therefore the phase-sum distribution is a useful
tool to probe the nonlinear effects.

The present analysis applies, for the first time, the above phase
statistics to the SDSS galaxy catalogs to quantify the non-Gaussianity in
galaxy distributions. There are three possible sources of the
non-Gaussianity in the present galaxy density field: primordial density
field, nonlinear gravitational evolution, and nonlinear galaxy biasing.
Recently, WMAP showed that the primordial density field is well
approximated by Gaussian statistics. Thus, the primordial non-Gaussianity
can be safely ignored in the following analysis, and the prominent source
of non-Gaussianity in the galaxy distribution is nonlinear gravitational
evolution. In our analysis, we use the phase statistics as a
cosmological tool independent of the two-point statistics. We compare
the phase correlations of SDSS galaxies with those of mock samples
constructed from $N$-body simulations based on various cosmological
models. For a fair comparison with observations, we consider the
observational systematic effects including the survey geometry, the
redshift distortion and the shot-noise due to the sparse sampling. We
find that the SDSS galaxy data agree very well with the LCDM model
predictions with $\sigma_8=0.9$, which is consistent with the results of
WMAP.

%%%%%%%%%%%%%%%%%%%%%%%%%%%%%%%%%%%%%%%%%%%%%%%%%%%%%%%
\begin{table*}[tph]
\begin{center}
\caption{Properties of our volume-limited samples $^*$.}
\begin{tabular}{cccccccccc}
\hline\hline
& & & & $V_{\rm samp}$ & & $l_{\rm mean}$ & $L_{\rm box}$ 
& Range of $k$ \\
\raisebox{1.5ex}[0pt]{$M_{r,{\rm min}}$} & 
\raisebox{1.5ex}[0pt]{$M_{r,{\rm max}}$} &
\raisebox{1.5ex}[0pt]{$z_{\rm min}$} &
\raisebox{1.5ex}[0pt]{$z_{\rm max}$} &
[$(h^{-1}{\rm Mpc})^3$] & \raisebox{1.5ex}[0pt]{$N_{\rm gal}$} & 
[$h^{-1}$Mpc] &[$h^{-1}$Mpc] & [$2\pi/(h^{-1}$Mpc)] \\ \hline
$-22.0$ & $-21.0$ & $0.067$ & $0.153$ & $2.92\times 10^7$ & $34008$ & $9.50$ & $800$ & $0.006$--$0.11$ \\
$-21.0$ & $-20.0$ & $0.044$ & $0.103$ & $9.20\times 10^6$ & $44636$ & $5.91$ & $560$ & $0.009$--$0.16$ \\
$-20.0$ & $-19.0$ & $0.028$ & $0.067$ & $2.69\times 10^6$ & $23099$ & $4.88$ & $360$ & $~~0.01$--$0.25$ \\
$-19.0$ & $-18.0$ & $0.018$ & $0.044$ & $7.46\times 10^5$ & $~8640$ & $4.42$ & $240$ & $~~0.02$--$0.37$ \\ \hline
\\
\multicolumn{9}{@{}l@{}}{\hbox to 0pt{\parbox{155mm}{\footnotesize
$^*$ Constructed from the SDSS galaxy catalog
`Large-scale Structure Sample 15' in Northern hemisphere. Listed
values are the upper and lower limits of the $r$-band magnitude,
$M_{r, {\rm min/max}}$, the upper and lower limits of the redshift,
$z_{\rm min/max}$, the survey volume, $V_{\rm samp}$, the total number
of galaxies, $N_{\rm gal}$, the mean separation of galaxies, $l_{\rm
mean}$, the box-size for Fourier transform, $L_{\rm box}$, and the
scale range of $k$ in Fourier space to be used in measuring the
distribution function of the phase sum.  LCDM model parameters,
$\Omega_{\rm m}=0.3$ and $\Omega_\Lambda=0.7$, are assumed throughout.}
\hss}}
\end{tabular}
\end{center}
\label{tab:vollim}
\end{table*}
%%%%%%%%%%%%%%%%%%%%%%%%%%%%%%%%%%%%%%%%%%%%%%%%%%%%%%%

Another possible source for non-Gaussianity, galaxy biasing, is the
uncertainty of the statistical relation between galaxies and the
underlying mass, which originates from complicated processes of galaxy
formation.  It is known that galaxy clustering depends sensitively on
various properties of galaxies, such as the luminosity, color,
morphology, and environment (e.g., \cite{Dressler1980}). The two-point
statistics using SDSS galaxies by \citet{Zehavi2005} and
\citet{Tegmark2004a} clearly exhibits the luminosity and morphology
dependence of the galaxy biasing (see also \cite{Kayo2004} for the
dependence in three-point correlation functions). If the density
fluctuation of galaxies, $\delta_{\rm g}(\mathbf x)$, is related to
the mass density fluctuation, $\delta_{\rm m}(\mathbf x)$, as
$\delta_{\rm g}=b_1\delta_{\rm m}$, their Fourier transforms also have
the linear relationship $\delta_{\rm g, \scriptsize\mathbf
k}=b_1\delta_{\rm m,\scriptsize\mathbf k}$. Decomposing the Fourier
wavevectors to the parts of amplitudes and phases, the amplitudes
change proportionally to $b_1$ regardless of the scale; however, the
phases are completely preserved under linear biasing. Therefore, the
agreement of the observed phase-sum distribution with the simulated
predictions suggests that the galaxy biasing is well approximated by
the linear relation if a correct cosmological model is assumed. For
definiteness, we adopt a quadratic deterministic biasing and put a
constraint on the degree of the nonlinearity of galaxy biasing in a
weakly nonlinear regime ($>30h^{-1}$Mpc).

This paper is organized as follows. In section 2 we describe the SDSS
data sets and our simulation mock catalogs.  Section 3 summarizes the
perturbative formula of phase-sum distribution by
\citet{Matsu2003} and explains the method of measuring phase
correlations using the distribution function of phase sum.  The
results of observed phase correlations are also presented. Section 4
is focused on the bispectrum analysis of $p^{(3)}$ from comparison
between observations and their mock catalogs. Finally section 5 is
devoted to a summary of our results and further discussion.

\section{Volume-Limited Samples for SDSS Galaxies and Mock Catalogs}
\label{sec:sample}

Our present analysis is based on a subset of the SDSS galaxy redshift
data, `Large-scale Structure Sample 15' \citep{Blanton2005}.  This
sample includes the spectroscopic data of $389306$ galaxies and covers
the sky area of $4426$ square degrees. The angular selection function
of the survey is written in terms of spherical polygons
\citep{Hamilton2004}.  Details of the SDSS can be found in the
following literature: \citet{York2000} provide an overview of the
SDSS.  Technical articles providing details of the SDSS include
descriptions of the photometric camera \citep{Gunn1998}, photometric
analysis \citep{Stoughton2002}, the photometric system and photometric
calibration \citep{Fukugita1996, Hogg2001, Ivezic2004, Smith2002}, the
photometric pipeline \citep{Lupton2001}, astrometric calibration
\citep{Pier2003}, selection of the galaxy spectroscopic samples
\citep{Eisen2001, Strauss2002}, and spectroscopic tiling
\citep{Blanton2003a}. The details of publicly released data are
summarized in \citet{Stoughton2002} for Early Data Release and
Abazajian et al. (2003, 2004, 2005) for Data Release One, Two and
Three, respectively.

In our analysis we use observational data in the Northern hemisphere,
except for the distant area from the other observed areas, which is
designed to overlap with the Spitzer Space Telescope First Look
Survey.  We also omit the area of three stripes in the Southern
hemisphere, which has a survey geometry not appropriate for the
analysis of galaxy clustering in three-dimensional space. The range of
the $r$-band apparent magnitude ($m_{r,{\rm min}}, m_{r,{\rm max}}$)
is set to be a conservative range of ($14.5, 17.5$) after correction
for Galactic reddening using the maps of \citet{SFD98}. We construct
four volume-limited samples with a unit width of the absolute magnitude,
which cover the range of the $r$-band absolute magnitude from $-22$ to
$-18$ (Table \ref{tab:vollim}). The redshift range of each
volume-limited sample ($z_{\rm min},z_{\rm max}$) is determined by the
following distance modulus relation:
%%%%%%%%%%%%%%%%%%%%%%%%%%%%
\begin{eqnarray}
M_{r,{\rm min/max}}&=&m_{r,{\rm min/max}} \nonumber \\
&-&5\log[r_{\rm min/max}(1+z_{\rm min/max})/10~{\rm pc}] \\
&-&K(z_{\rm min/max}), \nonumber
\end{eqnarray}
%%%%%%%%%%%%%%%%%%%%%%%%%%%%
where $r_{\rm min/max}$ is the comoving distance at a redshift of
$z_{\rm min/max}$ and $K(z)$ is a quadratic fitting formula of the
averaged $K$-correction as a function of $z$ \citep{Park2005},
%%%%%%%%%%%%%%%%%%%%%%%%%%%%%%%%%%%%%%%%%%%%%%%%%%%%%%%
\begin{equation}
K(z)=2.3537z^2+0.5735z-0.18437.
\end{equation}
%%%%%%%%%%%%%%%%%%%%%%%%%%%%%%%%%%%%%%%%%%%%%%%%%%%%%%%
The details of $K$-correction is described in \citet{Blanton2003b}.
Table \ref{tab:vollim} summarizes the 
properties of our volume-limited subsamples.

For a fair comparison with the observation, we construct a set of mock
simulation data, including observational effects of the survey
geometry, the number density, and the redshift distortion (Hikage et
al. 2002, 2003). For constructing mock samples, we use P${}^3$M
$N$-body simulations, provided by \citet{JS1998}. The simulation
employs $256^3$ particles in a periodic comoving box with a length of
$300h^{-1}{\rm Mpc}$ or $600h^{-1}$ Mpc using Gaussian initial
conditions and a Cold Dark Matter (CDM) transfer function
\citep{BBKS1986}.  We use the $z=0$ snapshot simulation data (for
simplicity we neglect the light-cone effect) in various CDM models
with the parameters listed in Table \ref{tab:modelpara}; the
dimensionless matter-density parameter, $\Omega_{\rm m}$, the
dimensionless cosmological constant, $\Omega_\Lambda$, the shape
parameter, $\Gamma$ of the CDM transfer function \citep{BBKS1986}, and
the r.m.s. density-fluctuation amplitude smoothed by a top-hat filter
with a scale of $8h^{-1}$Mpc, $\sigma_8$. We extract about $10$
realizations of wedge samples for each volume-limited sample out of
the full simulation cube so that they have the same sample-shape and
number of particles (averaged over mock samples) as each
volume-limited sample.  To construct mock samples that extend beyond
the simulation box size, we duplicate particles using the periodic
boundary conditions.  We create mock data in redshift space by adding
the line-of-sight component of the peculiar-velocity to the distance
of each particle for calculating redshift. The mock sample
simply assumes that each mass particle represents a simulated galaxy,
and neglects the effect of galaxy biasing, which are further
discussed in section \ref{sec:results}.

%%%%%%%%%%%%%%%%%%%%%%%%%%%%%%%%%%%%%%%%%%%%%%%%%%%%%%%%%%%%%%%%%%%
\begin{table}[tph]
\caption{Simulation model parameters}
\begin{center}
\begin{tabular}{ccccc}
\hline\hline
  Model & $\Omega_{\rm m}$ & $\Omega_\Lambda$ & $\Gamma$ 
  & $\sigma_8$ \\ \hline 
  LCDM & $0.3$ & $0.7$ & $0.21$ & $1$, $0.9$, $0.7$ \\ 
  SCDM & $1~~$ & $0~~$ & $0.5~$ & $0.6$ \\ 
  OCDM & $0.3$ & $0~~$ & $0.25$ & $1$ \\ \hline
\end{tabular}
\end{center}
\label{tab:modelpara}
\end{table}
%%%%%%%%%%%%%%%%%%%%%%%%%%%%%%%%%%%%%%%%%%%%%%%%%%%%%%%%%%%%%%%%%%%%%
\section{The Distribution Function of Phase Sum}
\label{sec:phasesum} 

\citet{Matsu2003} derived an analytical relation between the
polyspectra and the distribution of the phase sum in perturbation
theory. In the lowest-order approximation, the probability density
function (PDF) of the phase sum, $\theta_{{\scriptsize\mathbf
k}_1}+\theta_{{\scriptsize\mathbf k}_2} - \theta_{{{\scriptsize\mathbf
k}_1}+{{\scriptsize\mathbf k}_2}}$, over closed wavevectors ${\mathbf
k}_1, {\mathbf k}_2$, and $-{\mathbf k}_1-{\mathbf k}_2$ reduces to
\citep{Matsu2003}:
%%%%%%%%%%%%%%%%%%%%%%%%%%%%%%%%%%%%%%%%%%%%%%%%%%%%%%%
\begin{eqnarray}
\label{eq:matsubara}
&& {\cal P}(\theta_{{\scriptsize\mathbf
k}_1}+\theta_{{\scriptsize\mathbf k}_2} -\theta_{{{\scriptsize\mathbf
k}_1}+{{\scriptsize\mathbf k}_2}}|V_{\rm samp}) \cr &&  \propto 1+
\frac{\pi^{3/2}}{4}p^{(3)}({\mathbf k}_1,{\mathbf k}_2|V_{\rm samp})
\cos (\theta_{{\scriptsize\mathbf
k}_1}+\theta_{{\scriptsize\mathbf k}_2} -\theta_{{{\scriptsize\mathbf
k}_1}+{{\scriptsize\mathbf k}_2}}),
\end{eqnarray}
%%%%%%%%%%%%%%%%%%%%%%%%%%%%%%%%%%%%%%%%%%%%%%%%%%%%%%%
where $V_{\rm samp}$ is the sampling volume and $p^{(3)}$ is defined
by the bispectrum, $B({\mathbf k}_1,{\mathbf k}_2)$, and the power
spectrum, $P(k)$, as follows:
%%%%%%%%%%%%%%%%%%%%%%%%%%%%%%%%%%%%%%%%%%%%%%%%%%%%%%%
\begin{equation}
\label{eq:p3}
p^{(3)}({\mathbf k}_1,{\mathbf k}_2|V_{\rm samp})=\frac{B({\mathbf k}_1,
{\mathbf k}_2)}{\sqrt{V_{\rm samp}P(k_1)P(k_2)P(|{\mathbf k}_1+{\mathbf k}_2|)}}.
\end{equation}
%%%%%%%%%%%%%%%%%%%%%%%%%%%%%%%%%%%%%%%%%%%%%%%%%%%%%%%
In the lowest-order approximation, the PDF of the phase sum over three
wavevectors is fully determined by the parameter $p^{(3)}$.  If the
hierarchical clustering ansatz is valid, $B \sim P^2$ and thereby
$p^{(3)}$ is approximately given by $\sqrt{P/V_{\rm
samp}}$. Therefore, the lowest-order approximation [equation
(\ref{eq:matsubara})] breaks down at a scale $k$ satisfying the
condition $P(k)/V_{\rm samp} > 1$. Note that this is different from
the conventional condition of the gravitational nonlinear clustering,
$k^3 P(k) > 1$. In fact, a subsequent numerical analysis confirmed the
validity of the perturbative formula, even in a fairly nonlinear regime
of clustering \citep{Hikage2004}.

The dependence of the phase-sum distribution on the sampling volume
can be naturally explained as follows: the synchronization of phases
becomes weak when multiple density peaks with comparable heights exist
(section \ref{sec:intro}). Because the sampling volume is larger, the
number of peaks increases, and thus the phase-sum distribution
approaches being uniform.

When $V_{\rm samp}$ is sufficiently large, the higher order terms
above $p^{(3)}$ become negligible and then equation
(\ref{eq:matsubara}) is applicable, even in nonlinear regimes, in the
sense that $k^3 P(k)>1$.  On the contrary, decreasing $V_{\rm samp}$
by dividing the original survey volume into several sub-volumes would
provide information about the higher-order spectrum. Dividing the
original sample, however, leads to decreasing the number of objects
per sub-sample. Thus, a significant fraction of the available Fourier
modes suffers from the contamination of shot noise. In the present
galaxy sample, it is quite difficult to extract higher order
information beyond $p^{(3)}$ from the distribution function of the
three-point phase sum. In the present work, therefore, we do not
divide the survey volume of each volume-limited sample.

We compute the distribution function of the phase sum for density
contrast fields from mock samples and SDSS galaxies.  We use the
cloud-in-cell interpolation to assign galaxies or dark matter
particles to mesh densities. Then, the density contrast field is
Fourier-transformed on cubic grids with a mesh number $N_{\rm
mesh}=256$ per side. We calculate the sum of the Fourier phases for
the closed set of three wavevectors, and then compute the distribution
function of the phase sum within each binning range of triangle
configurations.  For simplicity, we focus on the configuration of
wavevectors to be a nearly isosceles triangle, where the absolute
values of two wavevectors, $k_1$ and $k_2$, are within the same
binning range of scale $k$. The scale $k$ is divided into $8$ bins
with equal width in logarithmic scale in the range of $a<k/k_{\rm
Nyq}<b$. The Nyquist frequency $k_{\rm Nyq}$ is equal to $2\pi\times
(N_{\rm mesh}/2)/L_{\rm box}$, where $L_{\rm box}$ is a side of a
cubic box for a Fourier transformation.  Table \ref{tab:vollim} lists
$L_{\rm box}$ for each volume-limited sample, which is chosen for the
sample volume to be barely covered. We limit the range of scale to be
$a=0.04$ and $b=0.7$ so that the observational systematics due to the
complicated survey geometry and the discreteness effect may be
neglected compared to the sample variance; at smaller $k$, the Poisson
error increases due to the limited number of independent combinations
of modes forming closed triangles, and furthermore the convolution
effect with the survey mask is more serious. At larger $k$, the shot
noise becomes more serious. Table \ref{tab:vollim} lists the range of
scale $k$ for each volume-limited sample.  The unit of scale $k$ in
Fourier space is set to be $2\pi/(h^{-1}{\rm Mpc})$, and thereby the
corresponding scale-length in real space is just the inverse of $k$ in
units of $h^{-1}$Mpc. The angle $\varphi$ between ${\mathbf k}_1$ and
${\mathbf k}_2$, i.e., $\varphi\equiv\arccos[({\mathbf
k}_1/k_1)\cdot({\mathbf k}_2/k_2)]$ with $0^\circ<\varphi<180^\circ$, is
binned to have an equal width of $20^\circ$.

%%%%%%%%%%%%%%%%%%%%%%%%%%%%%%%%%%%%%%%%%%%%%%%%%%%%%%%
\begin{figure*}[tph]
\centering \FigureFile(80mm,80mm){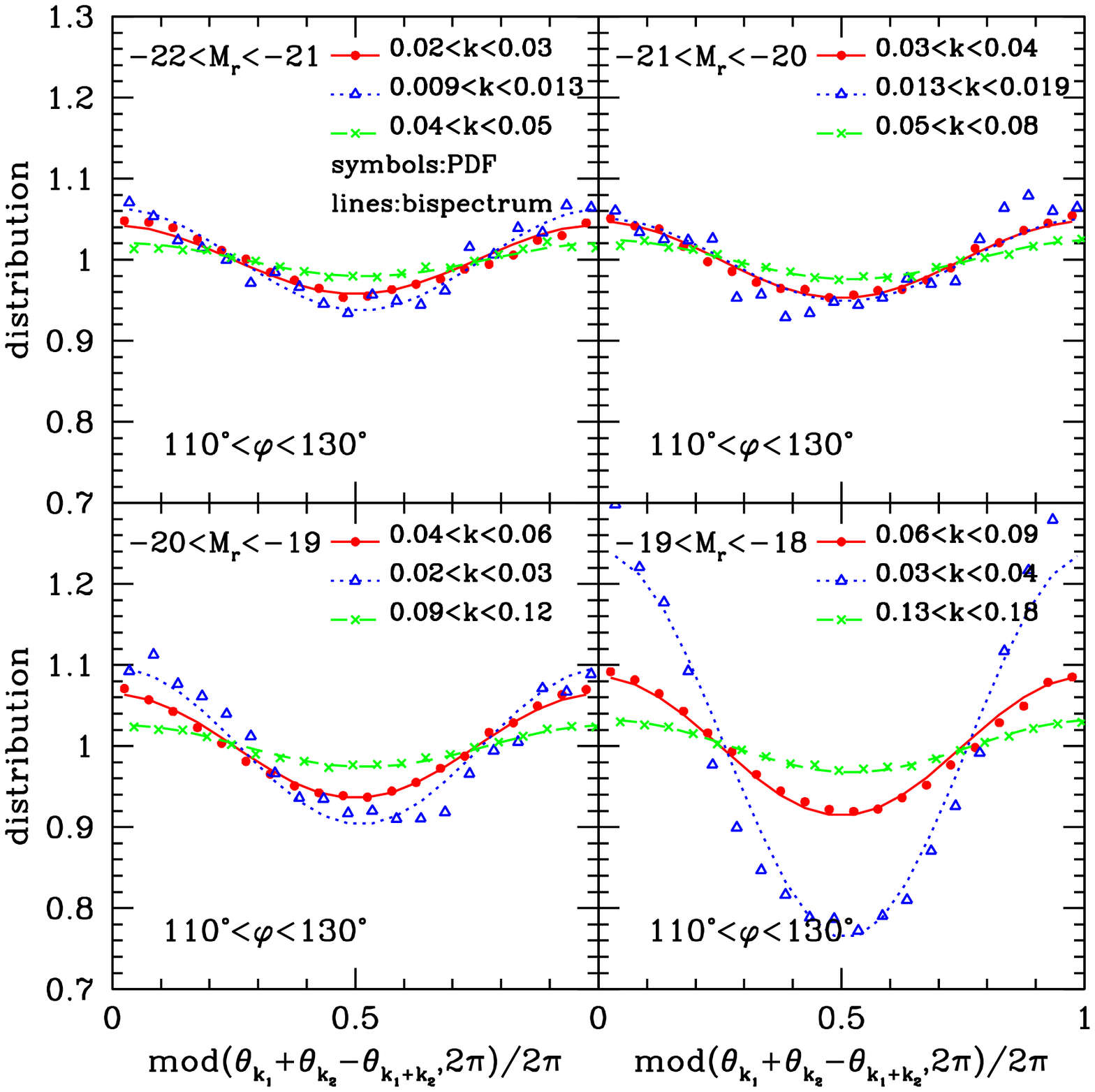}
\centering \FigureFile(80mm,80mm){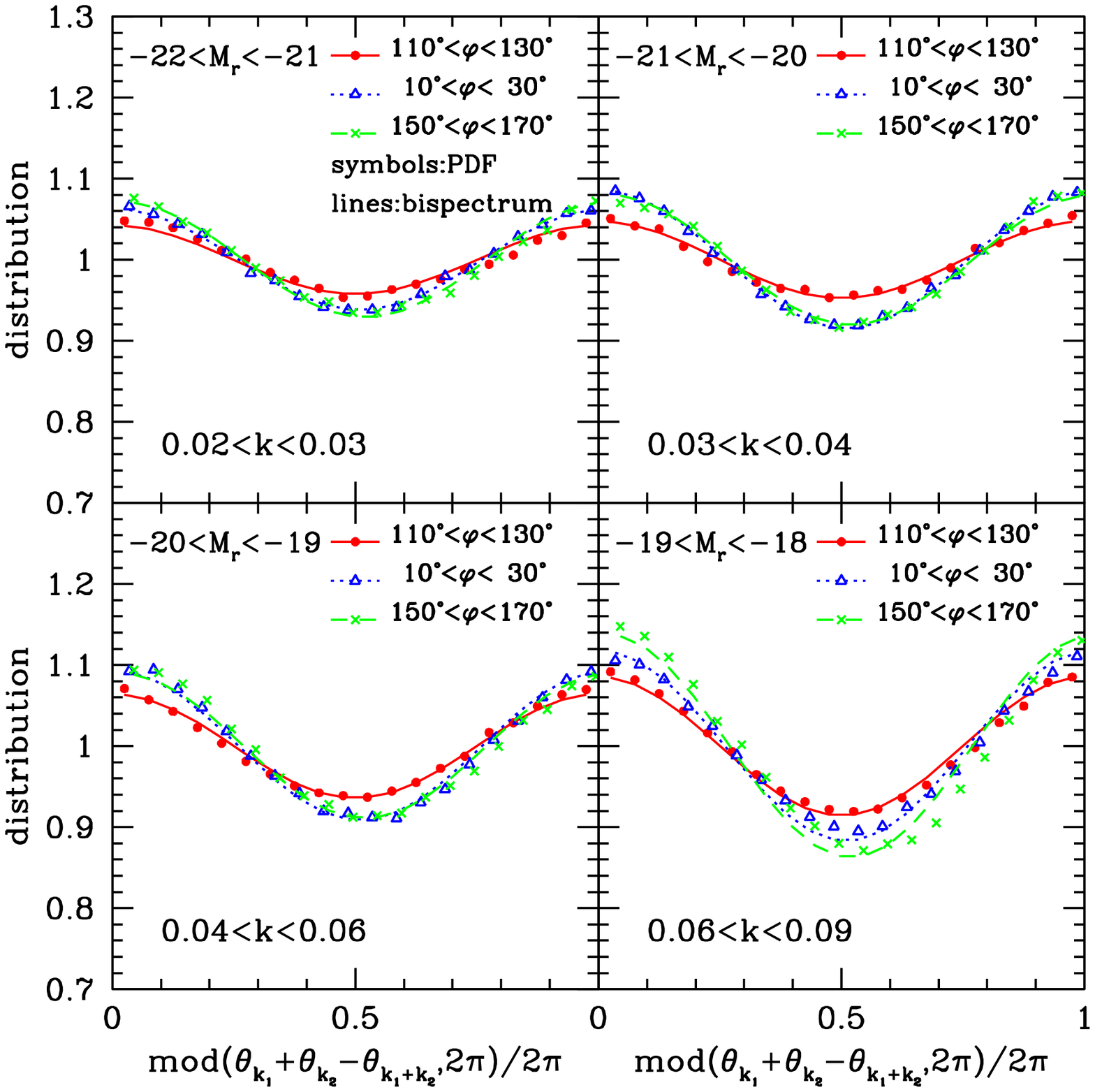}
\caption{Comparison of the observed distribution function of the phase
sum (symbols) with the lowest-order approximation [equation
($\ref{eq:matsubara}$)] where $p^{(3)}$ is computed by the combination
of the bispectrum and the power spectrum from observations [equation
($\ref{eq:p3}$)] (lines).  Different panels show the results for
different volume-limited samples.  In left figure, the configurations
of triangle wavevectors are focused on nearly equilateral triangle
($110^\circ<\varphi<130^\circ$) with different scales of $k\sim
k_1\sim k_2$. In right figure, the triangle configurations are nearly
isosceles ($k\sim k_1\sim k_2$) with different angles $\varphi$
between $k_1$ and $k_2$. The unit of scale $k$ is set to be
$2\pi/(h^{-1}{\rm Mpc})$.}
\label{fig:obsphasesum}
\end{figure*}
%%%%%%%%%%%%%%%%%%%%%%%%%%%%%%%%%%%%%%%%%%%%%%%%%%%%%%%

Figure \ref{fig:obsphasesum} plots the distribution function of the
phase sum for each volume-limited sample in symbols. For comparison,
we plot in lines the lowest-order approximation [equation
(\ref{eq:matsubara})] using $p^{(3)}$ computed from the bispectrum and
the power spectrum (\ref{eq:p3}).  In the left panels, the triangle
configurations are nearly equilateral shape
($110^\circ<\varphi<130^\circ$) for different $k$. In the right
panels, the triangle configurations are nearly isosceles for different
angles, $\varphi$. The phase-sum distributions are found to be well
approximated by the lowest-order approximation for all of the
volume-limited samples, regardless of the triangle configurations.
The dependence of $p^{(3)}$ on $V_{\rm samp}$ and $k$ can be
understood from the nature of perturbative parameter, $p^{(3)}$, which
is roughly proportional to $[P(k)/V_{\rm samp}]^{1/2}$ under the
hierarchical clustering ansatz; volume-limited samples for fainter
galaxies have a smaller $V_{\rm samp}$ (increase $p^{(3)}$), but a
smaller $P(k)$ (decrease $p^{(3)}$) because the measurable range of
scale $k$ is shifted to larger values due to the smaller $V_{\rm
box}$. The redshift distortion due to the random motion of galaxies
also smears the power at small scales, and thus $p^{(3)}$ reaches up to
$0.3$ at a maximum, much smaller than unity. Figure
\ref{fig:obsphasesum} shows that $p^{(3)}$ calculated from the
amplitude of the phase-sum distribution using equation
(\ref{eq:matsubara}) is nearly equal to that from the combination of
the bispectrum and the power spectrum [equation (\ref{eq:p3})].  In
what follows, we will use equation (\ref{eq:p3}) to compute $p^{(3)}$
instead of fitting the PDF to equation (\ref{eq:matsubara}).

\section{Bispectrum Analysis of $p^{(3)}$}
\label{sec:results}

\begin{figure*}[tph]
\centering \FigureFile(80mm,80mm){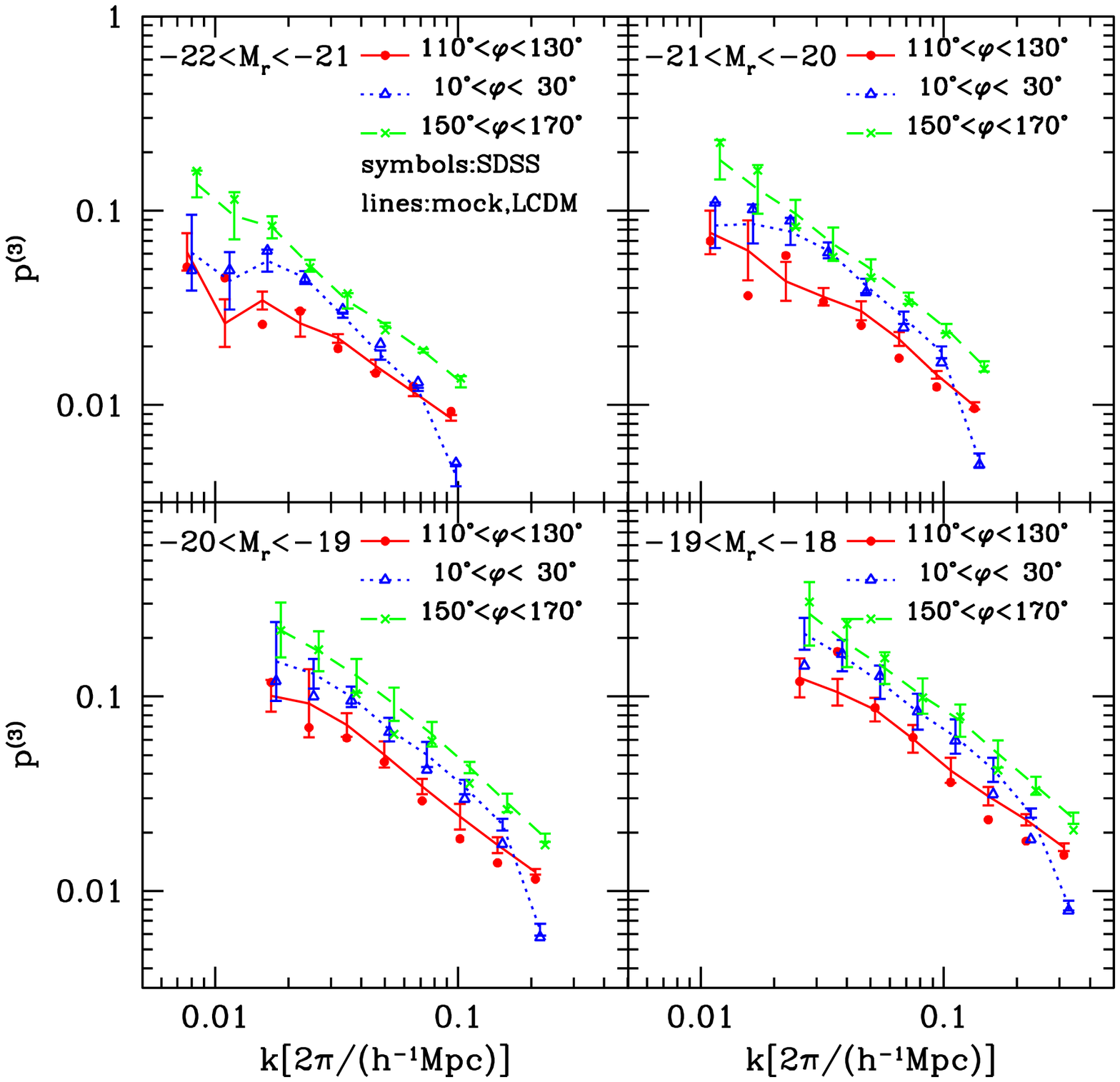}
\centering \FigureFile(80mm,80mm){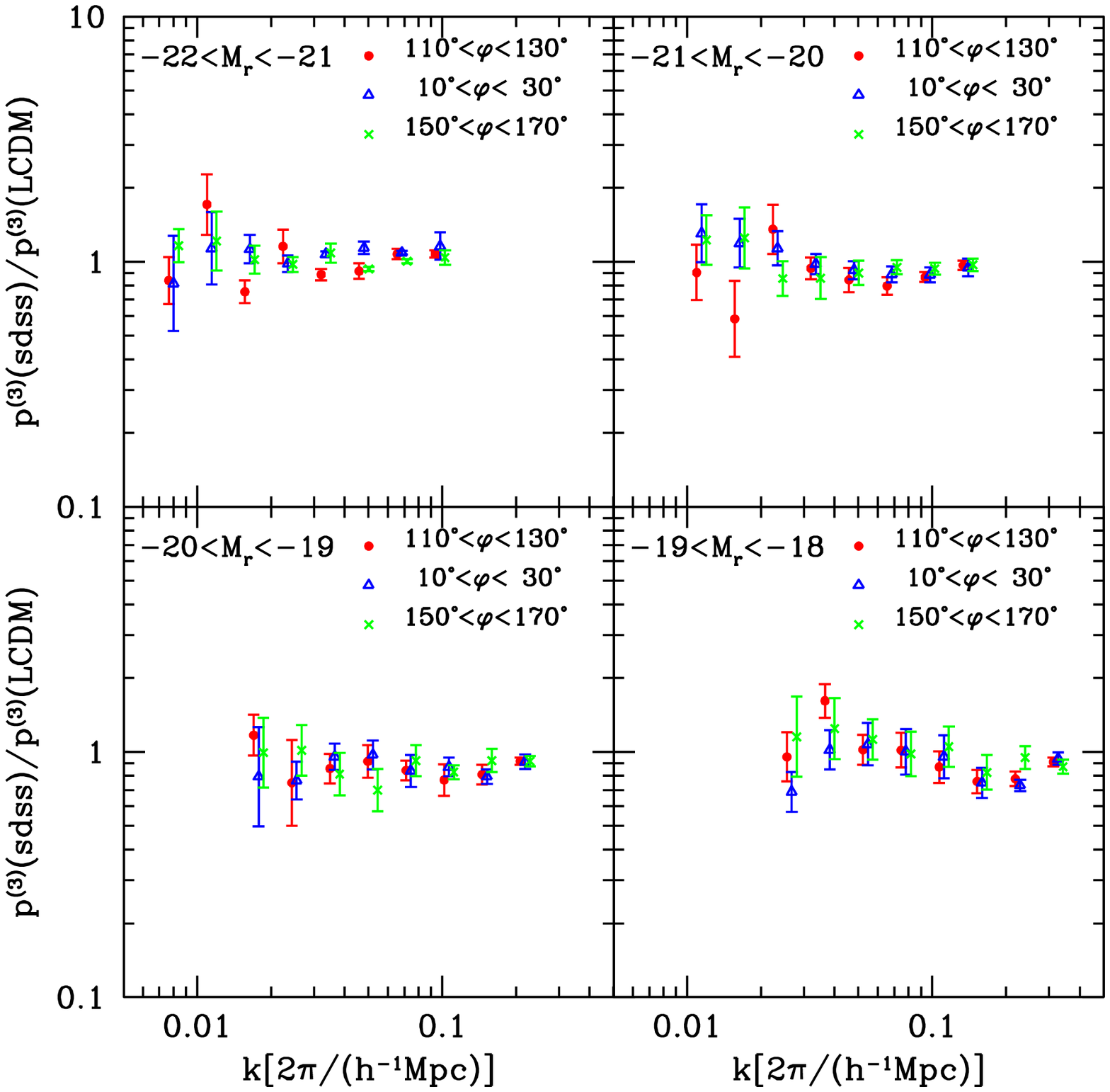}
\caption{Scale dependence of the observed $p^{(3)}$ for each
volume-limited sample (symbols). For comparison, the estimations of
mock samples based on the representative model of LCDM with
$\sigma_8=0.9$ are plotted by lines. The right figures show the ratio of
the observed $p^{(3)}$ and the corresponding mock estimations.  The
configurations of triangle wavevectors are equilateral triangles with
different scales of $k$. Error-bars represent the sample
variance of the mock samples for each sample.}
\label{fig:obsp3_scale}
\end{figure*}

\begin{figure*}[tph]
\centering \FigureFile(80mm,80mm){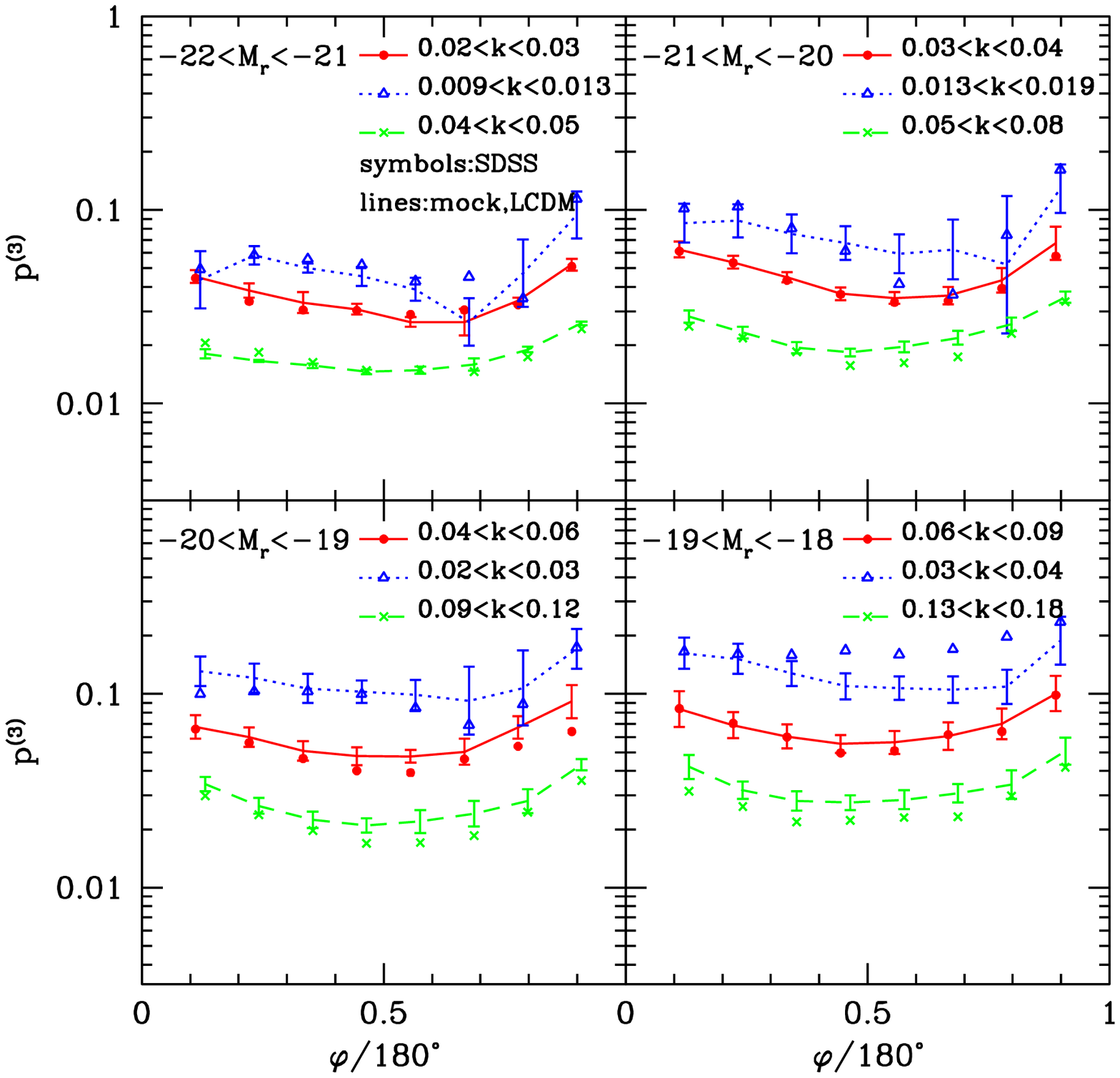}
\centering \FigureFile(80mm,80mm){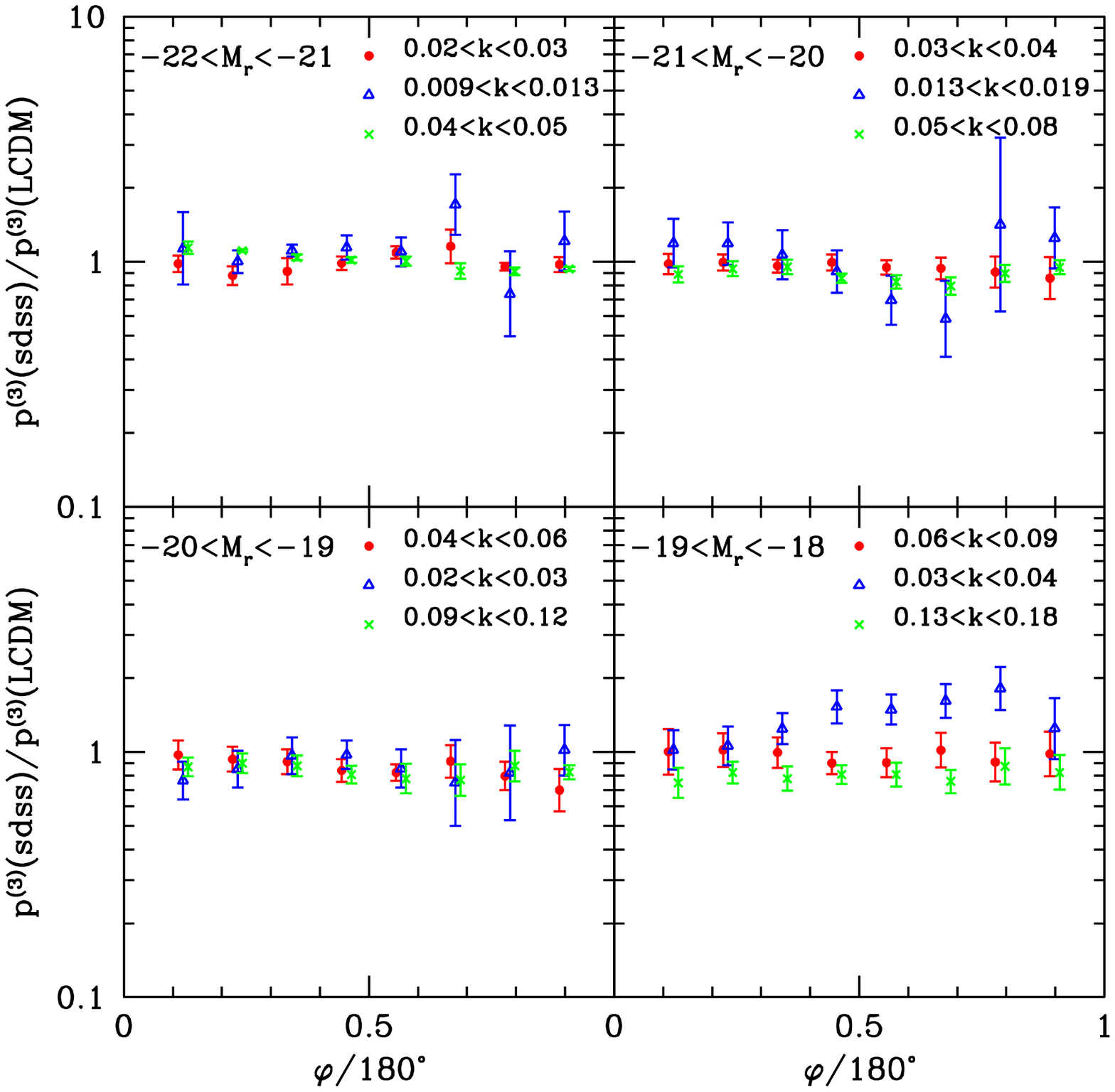}
\caption{Same as Figure \ref{fig:obsp3_scale}, but for the angle $\varphi$
dependence. The configurations of triangle wavevectors are isosceles
with different $\varphi$.}
\label{fig:obsp3_shape}
\end{figure*}

Figures \ref{fig:obsp3_scale} and $\ref{fig:obsp3_shape}$ show 
$p^{(3)}$ for SDSS galaxies in different volume-limited samples
(plotted in symbols) against the wavenumber $k$ (Figure
\ref{fig:obsp3_scale}) and the angle $\varphi$ (Figure
\ref{fig:obsp3_shape}). For comparison, the LCDM predictions with
$\sigma_8=0.9$ are plotted in lines.  Clearly, the scale-dependence of
$p^{(3)}$ is well approximated by a power-law of $k$. This can be
understood again from equation (\ref{eq:p3}); because $B(k_1,k_2)$ and
$P(k)$ may be approximated as power-laws at those scales, we expect
$p^{(3)} \propto \sqrt{P(k)} \propto k^{-1}$.  The right panels show the
ratio of the observed $p^{(3)}$ to the corresponding mock estimations
in the left panels. The overall agreement between observations and
LCDM predictions with $\sigma_8=0.9$ is very good.

\begin{figure*}[tph]
\centering \FigureFile(80mm,80mm){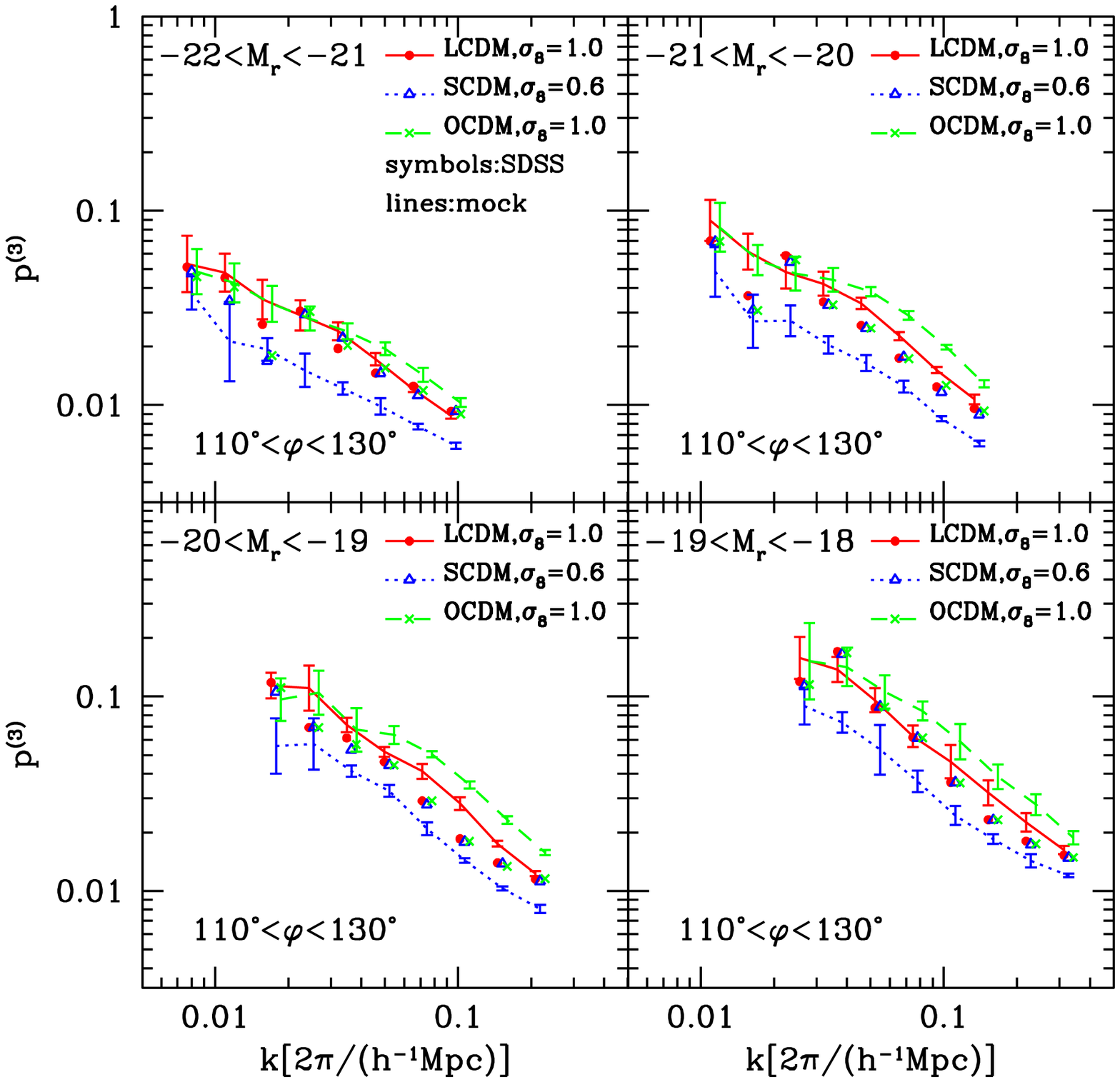}
\centering \FigureFile(80mm,80mm){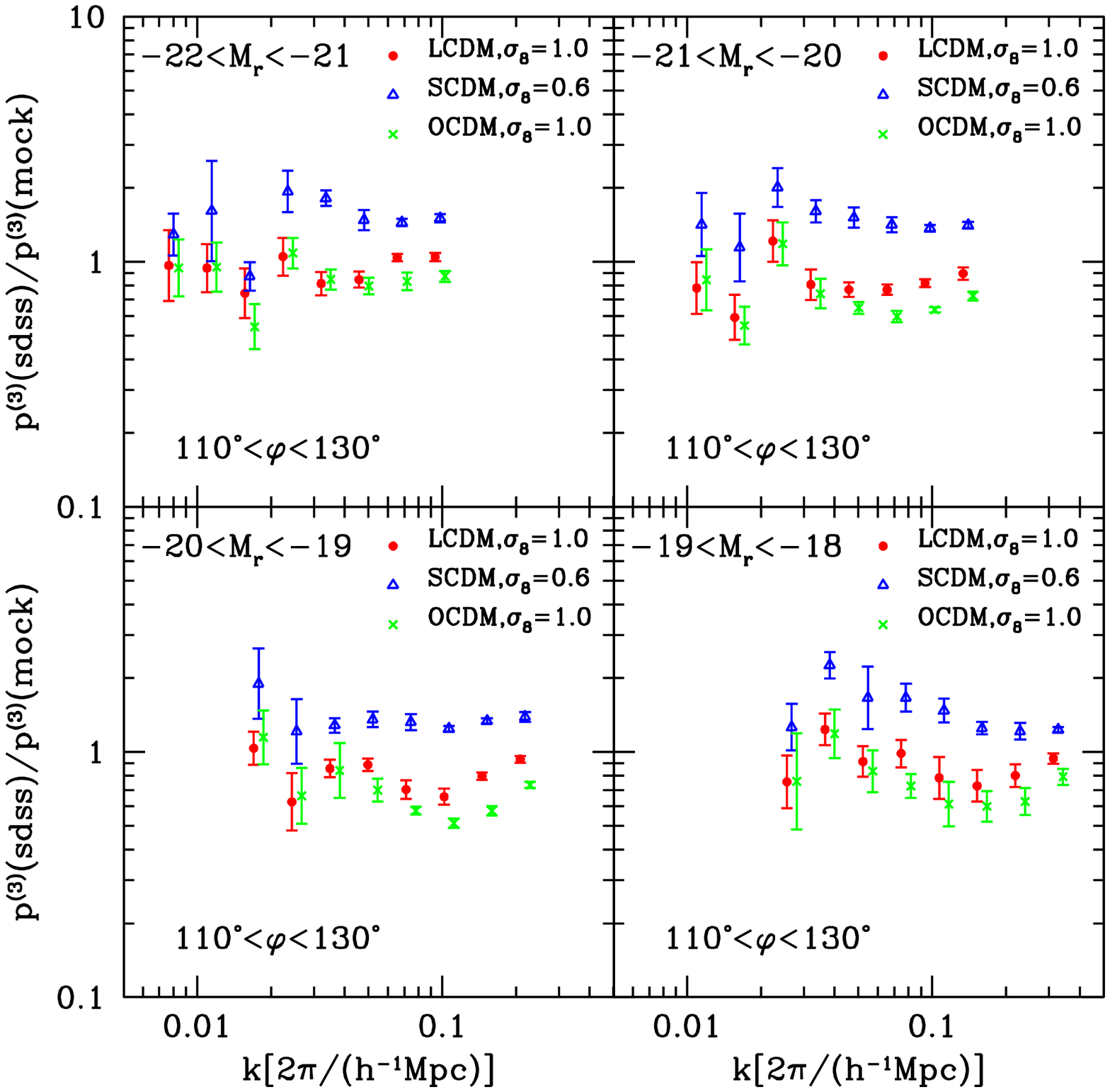}
\caption{Comparison of the observed $p^{(3)}$ (symbols) with the
simulated estimations (lines) based on various cosmologies, in
particular focused on the scale dependence; LCDM with $\sigma_8=1$
(filled circles and solid lines), SCDM (open triangles and dotted
lines), and OCDM (crosses and dashed lines). The right figures show the
ratio of the observed $p^{(3)}$ to the simulated $p^{(3)}$ for each
cosmology corresponding to the left figures. The error-bars represent the
sample variance of the mock samples. The configuration formed by three
wavevectors is in the shape of nearly equilateral triangles
($110^\circ<\varphi<130^\circ$).}
\label{fig:obsp3_scale_cdm}
\end{figure*}

\begin{figure*}[tph]
\centering \FigureFile(80mm,80mm){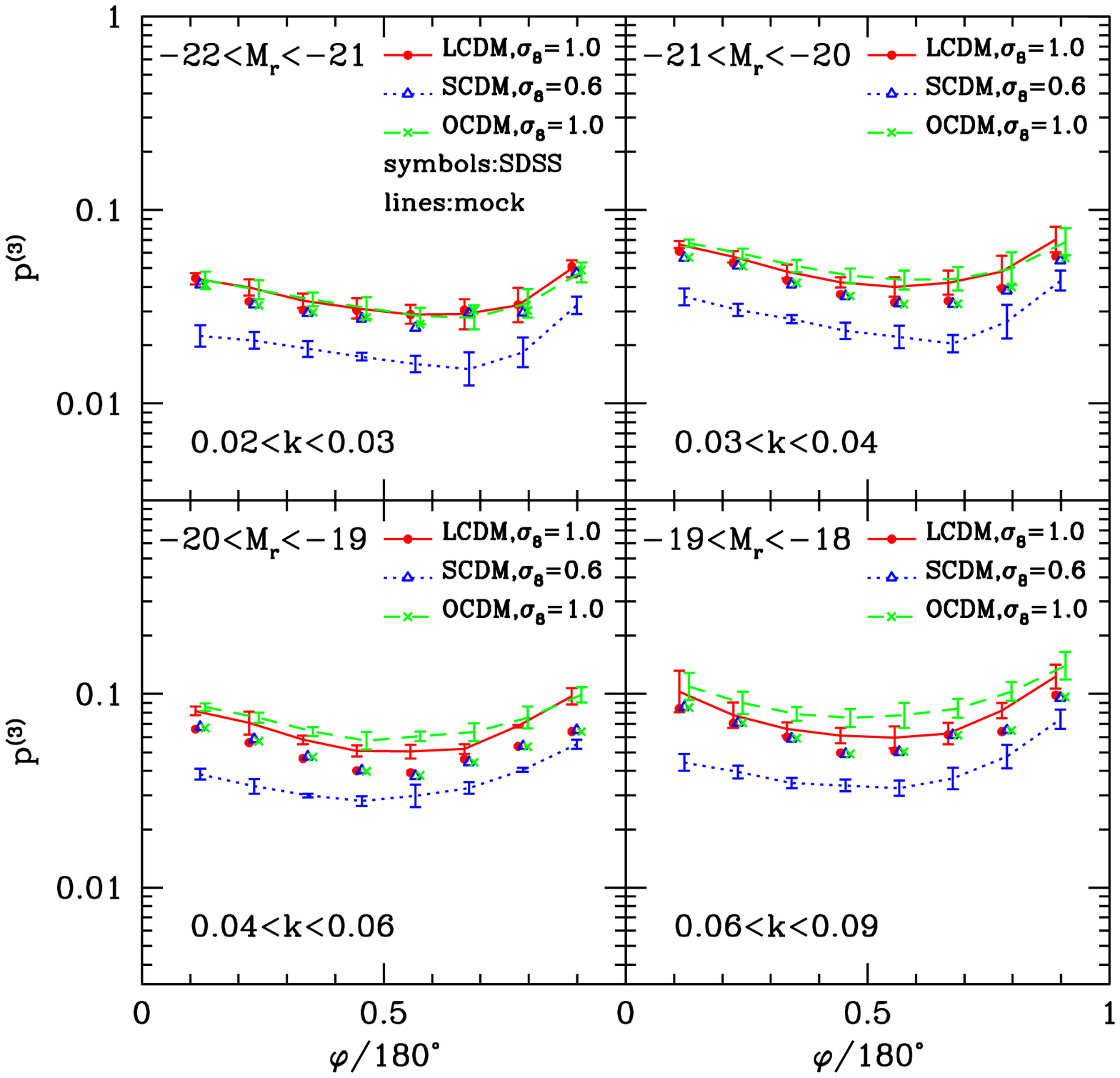}
\centering \FigureFile(80mm,80mm){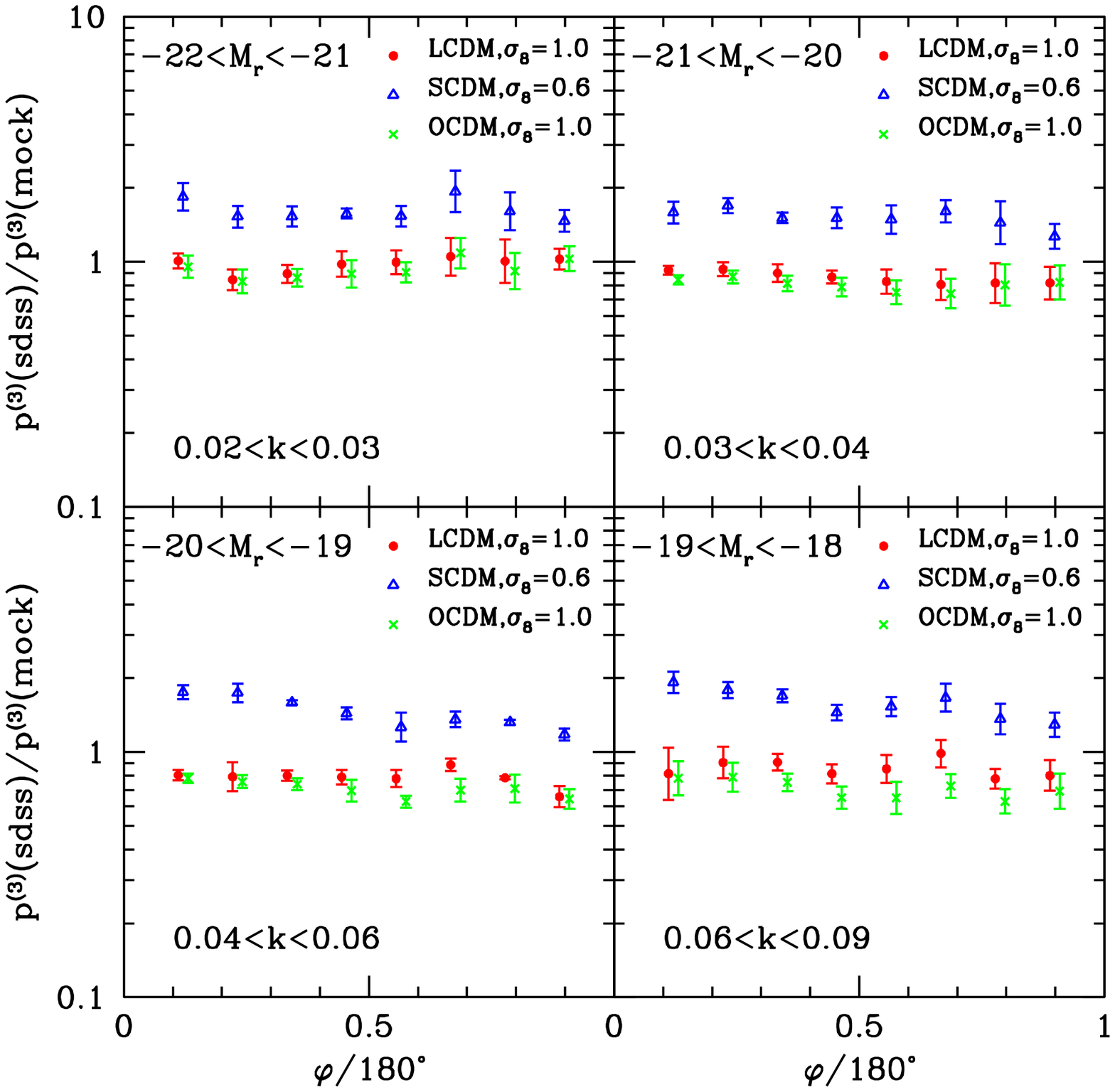}
\caption{Same as Figure \ref{fig:obsp3_scale_cdm}, but for angle $\varphi$
dependence. The configuration of the triangle is isosceles ($k_1\sim
k_2\sim k$) and thus the plotted scale is within the intermediate
bin of scales for each volume-limited sample.}
\label{fig:obsp3_shape_cdm}
\end{figure*}

\begin{figure*}[tph]
\centering \FigureFile(80mm,80mm){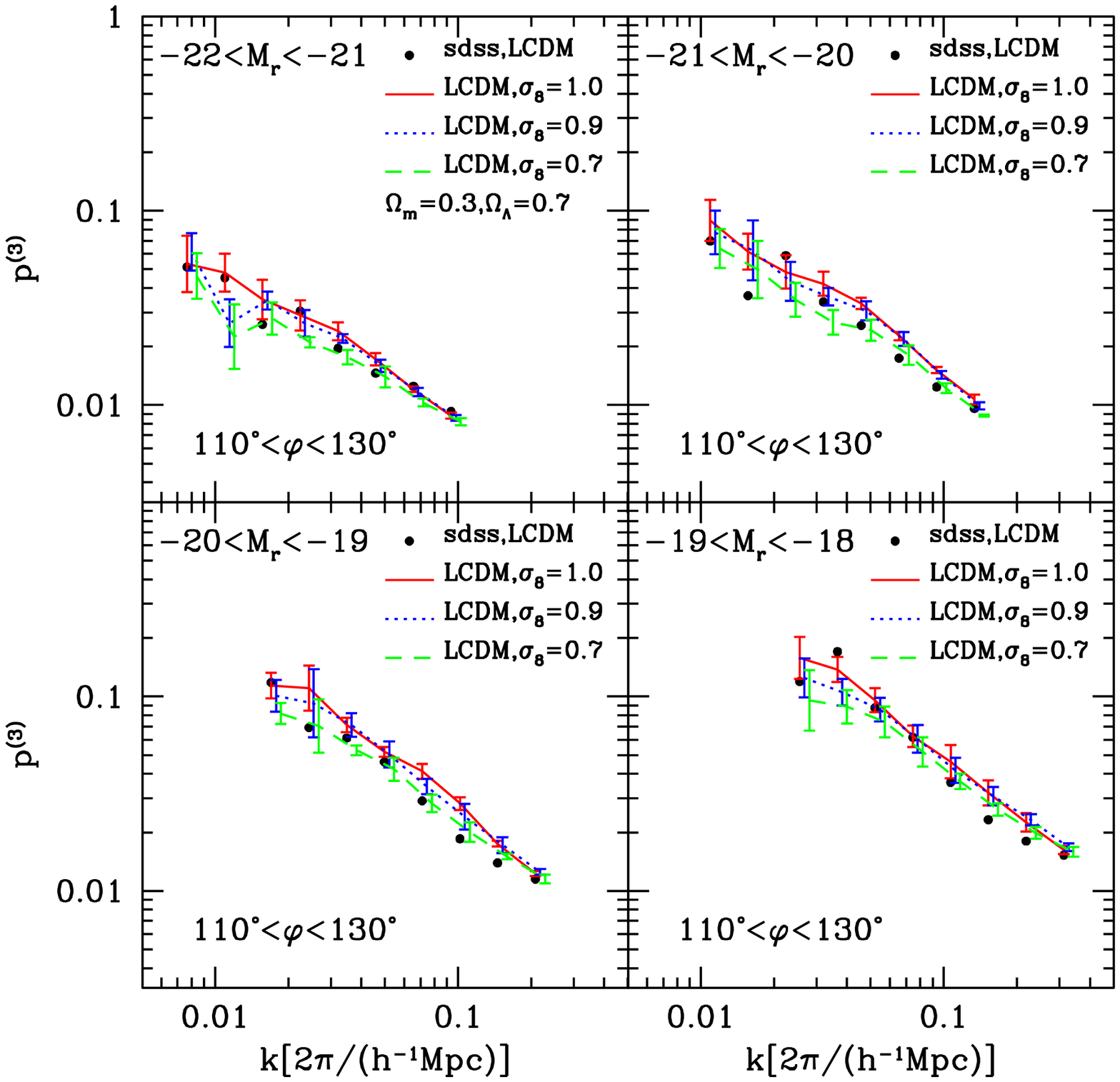}
\centering \FigureFile(80mm,80mm){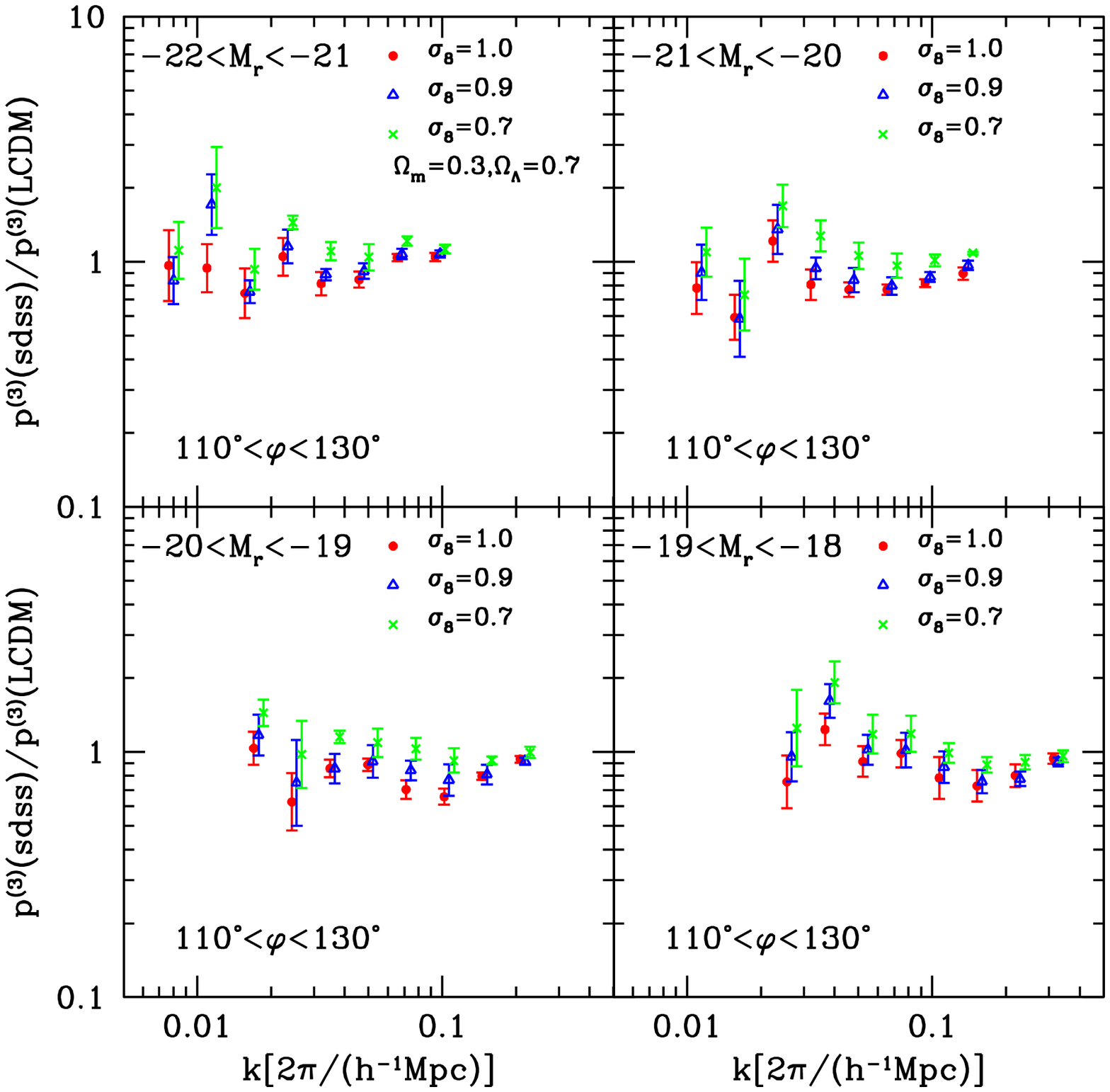}
\caption{Same as Figure \ref{fig:obsp3_scale_cdm}, but for compared
mock samples based on LCDM models with $\sigma_8=1$ (filled circles and
solid lines), $\sigma_8=0.9$ (open triangles and dotted lines), and
$\sigma_8=0.7$ (crosses and dashed lines).}
\label{fig:obsp3_scale_sig8}
\end{figure*}

\begin{figure*}[tph]
\centering \FigureFile(80mm,80mm){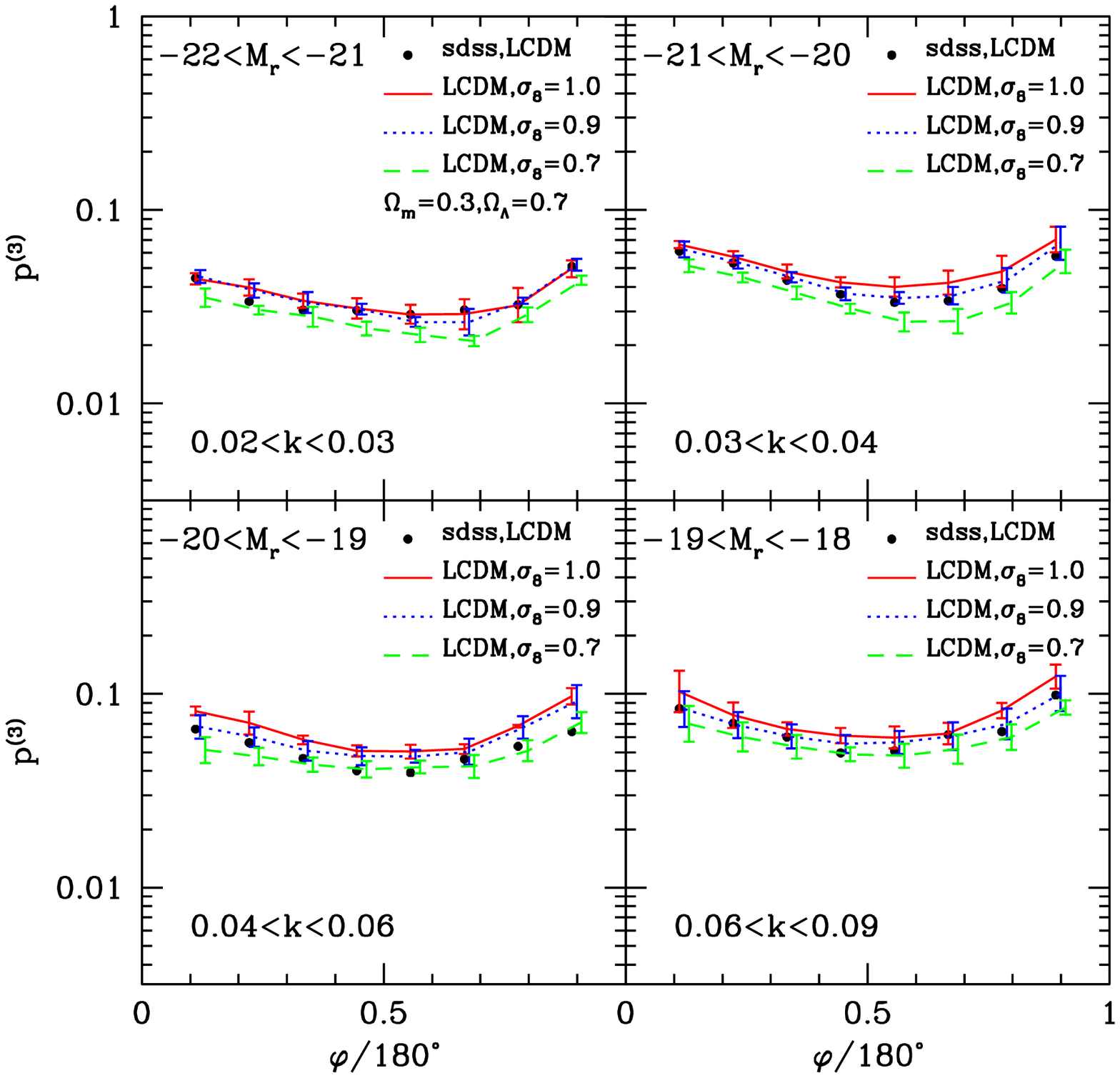}
\centering \FigureFile(80mm,80mm){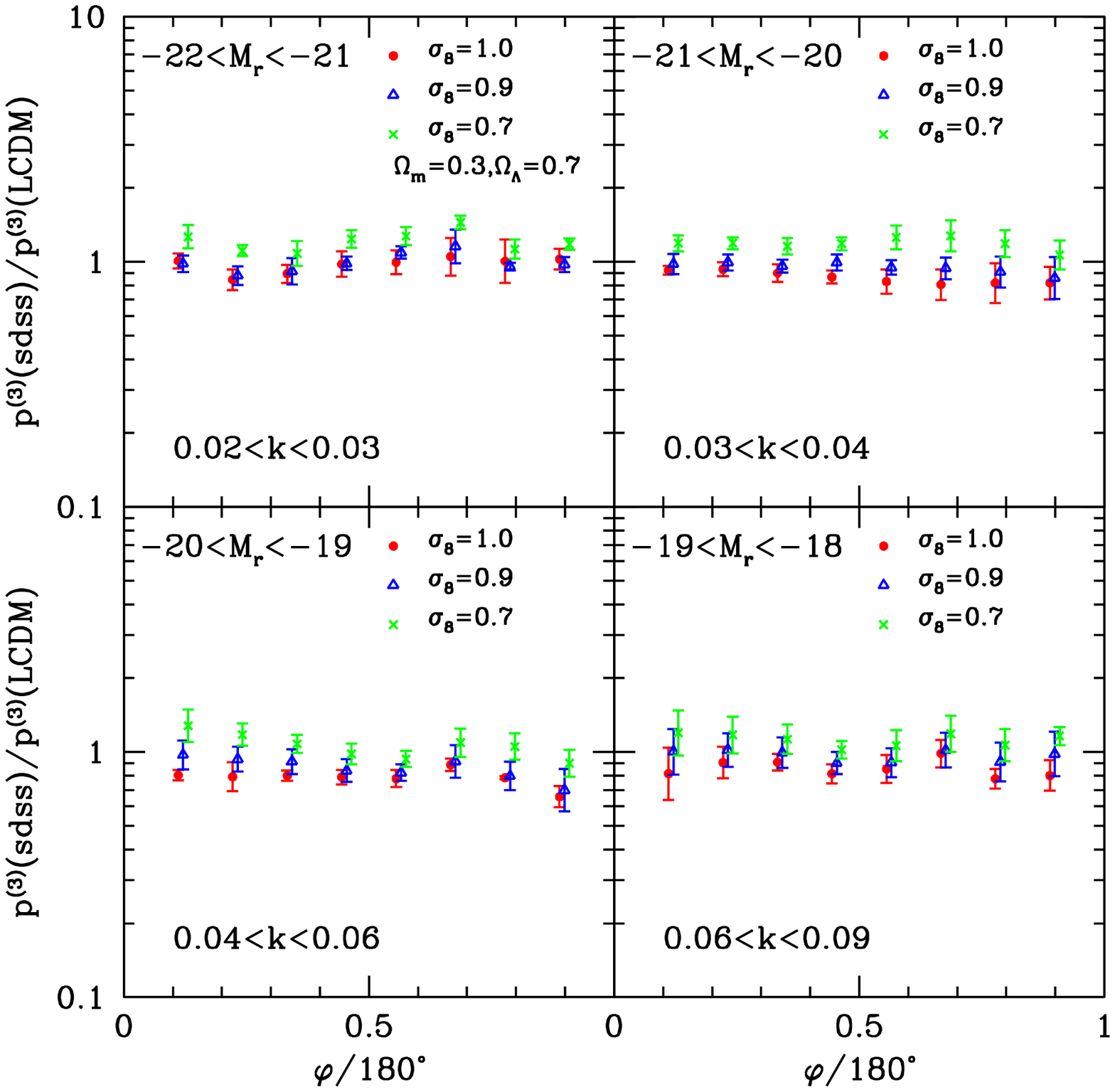}
\caption{Same as Figure \ref{fig:obsp3_scale_sig8}, but for the angle
$\varphi$ dependence. The configuration of the triangle is isosceles
($k_1\sim k_2\sim k$), and thus the plotted scale is within the
intermediate bin of scales for each volume-limited sample.}
\label{fig:obsp3_shape_sig8}
\end{figure*}

Figures \ref{fig:obsp3_scale_cdm} and \ref{fig:obsp3_shape_cdm} show
comparisons of the observed $p^{(3)}$ with various mock predictions to
examine the cosmology dependence. Figure \ref{fig:obsp3_scale_cdm}
focuses on the scale dependence, while Figure
\ref{fig:obsp3_shape_cdm} focuses on the shape dependence (the angle
$\varphi$ of the triangles).  The right panels show the ratio of
$p^{(3)}$ between observations and the corresponding mock samples in
the left panels. Again, LCDM predictions with $\sigma_8=0.9$ show the
best agreement with the observations among our mock samples; the SCDM
model predicts a smaller $p^{(3)}$ than the other models over the
whole range of scales, mainly because of the smaller value of
$\sigma_8$ (see Table \ref{tab:modelpara}). The OCDM model has the
same value of $\sigma_8$ as the LCDM models, but their spectral shape
is different: the small-scale fluctuation in OCDM is slightly larger
due to the larger value of $\Gamma$, and thus OCDM predicts higher
values of phase correlations than LCDM.  Figures
\ref{fig:obsp3_scale_sig8} and \ref{fig:obsp3_shape_sig8} are same as
Figures \ref{fig:obsp3_scale_cdm} and \ref{fig:obsp3_shape_cdm}, but
for the $\sigma_8$ dependence using mock samples based on the same
cosmology of LCDM. A systematic increase of $p^{(3)}$ as $\sigma_8$ is
clearly found because the phase correlations become strong as the
gravitational evolution proceeds.

%%%%%%%%%%%%%%%%%%%%%%%%%%%%%%%%%%%%%%%%%%%%%%%%%%%%%%%
%%%%%%%%%%%%%%%%%%%%%%%%%%%%%%%%%%%%%%%%%%%%%%%%%%%%%%%
%%%%%%%%%%%%%%%%%%%%%%%%%%%%%%%%%%%%%%%%%%%%%%%%%%%%%%%

%%%%%%%%%%%%%%%%%%%%%%%%%%%%%%%%%%%%%%%%%%%%%%%%%%%%%%%
%%%%%%%%%%%%%%%%%%%%%%%%%%%%%%%%%%%%%%%%%%%%%%%%%%%%%%%

%%%%%%%%%%%%%%%%%%%%%%%%%%%%%%%%%%%%%%%%%%%%%%%%%%%%%%%
%%%%%%%%%%%%%%%%%%%%%%%%%%%%%%%%%%%%%%%%%%%%%%%%%%%%%%%

%%%%%%%%%%%%%%%%%%%%%%%%%%%%%%%%%%%%%%%%%%%%%%%%%%%%%%%
%%%%%%%%%%%%%%%%%%%%%%%%%%%%%%%%%%%%%%%%%%%%%%%%%%%%%%%

%%%%%%%%%%%%%%%%%%%%%%%%%%%%%%%%%%%%%%%%%%%%%%%%%%%%%%%
%%%%%%%%%%%%%%%%%%%%%%%%%%%%%%%%%%%%%%%%%%%%%%%%%%%%%%%

Galaxy biasing is another source of uncertainty in comparing the
observations and the mock predictions. If the galaxy biasing is
linear, i.e., $\delta_{\rm g}=b_1\delta_{\rm m}$, the amplitude
changes in proportion to the linear coefficient, $b_1$, but the phases
are invariant. Thus, the consistency of the LCDM model remains valid
as long as the galaxy biasing is well approximated by the linear
relation in {\it redshift} space (note that this does not strictly
correspond to the linear biasing in {\it real} space).  The analysis
of the bispectrum for 2dFGRS finds that galaxy biasing is
consistent to be linear under LCDM model in the range of scales from
$5h^{-1}$Mpc to $30h^{-1}$Mpc \citep{Verde2002}.

Indeed, the combined analysis of three-point correlation functions and
two-point correlation functions shows a significant deviation from
linear biasing at {\it nonlinear} scales from $1h^{-1}$Mpc to
$10h^{-1}$Mpc \citep{Kayo2004}.  In contrast, the scales that we probe
in the present analysis is much larger, i.e., $> 30h^{-1}$Mpc.  We
therefore parameterize the galaxy biasing by the local and quadratic
relation between the density fluctuation of galaxies, $\delta_{\rm g}$,
and that of mass, $\delta_{\rm m}$, as follows;
%%%%%%%%%%%%%%%%%%%%%%%%%%%%%%%%%%%%%%%%%%%%%%%%%%%%%%%
\begin{equation}
\delta_{\rm g}=b_1\delta_{\rm m}+\frac{b_2}{2}(\delta_{\rm m}^2-
\langle\delta_{\rm m}^2\rangle),
\end{equation}
%%%%%%%%%%%%%%%%%%%%%%%%%%%%%%%%%%%%%%%%%%%%%%%%%%%%%%%
where $b_1$ and $b_2$ are biasing parameters. In a weakly nonlinear
regime, for example at scales larger than $30h^{-1}$Mpc, the locality of
the galaxy biasing is valid because the scale of galaxy formation is
much smaller ($\sim$ Mpc). Also the density fluctuation is less than
unity, and thereby higher order terms of the density fluctuation are
negligible. Under the above quadratic biasing,  $p_{\rm g}^{(3)}$ for
galaxies and $p_{\rm m}^{(3)}$ for mass are related as
%%%%%%%%%%%%%%%%%%%%%%%%%%%%%%%%%%%%%%%%%%%%%%%%%%%%%%%
\begin{equation}
p^{(3)}_{\rm g} ({\mathbf k}_1,{\mathbf k}_2)=p^{(3)}_{\rm m} ({\mathbf k}_1,{\mathbf k}_2)
+\frac{b_2}{b_1}f(P_1,P_2,P_3),
\label{eq:p3bias}
\end{equation}
%%%%%%%%%%%%%%%%%%%%%%%%%%%%%%%%%%%%%%%%%%%%%%%%%%%%%%%
\begin{equation}
f(P_1,P_2,P_3)=\frac{P_1P_2+P_2P_3+P_3P_1}{\sqrt{V_{\rm samp}
P_1P_2P_3}},
\label{eq:p3biasfactor}
\end{equation}
%%%%%%%%%%%%%%%%%%%%%%%%%%%%%%%%%%%%%%%%%%%%%%%%%%%%%%%
where $P_1=P(k_1), P_2=P(k_2)$, and $P_3=P(|{\mathbf k}_1+{\mathbf
k}_2|)$.  Equation (\ref{eq:p3bias}) implies that the difference of
$p^{(3)}_{\rm g}$ from $p^{(3)}_{\rm m}$ is proportional to the ratio
$b_2/b_1$ alone.

\begin{figure}[tph]
\centering \FigureFile(80mm,80mm){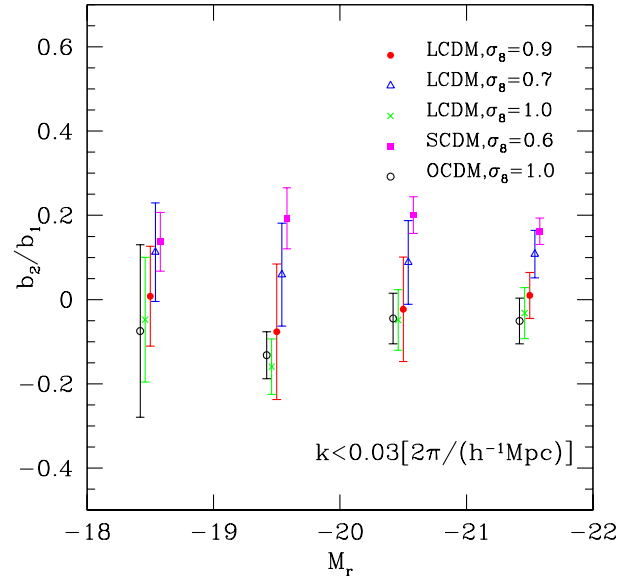}
\caption{Averaged ratio of the quadratic bias parameter to the linear
bias parameter, $b_2/b_1$, estimated from the difference of $p^{(3)}$
between observations and mock predictions using equation
(\ref{eq:p3bias}).  Cosmological models used in mock predictions
are LCDM with $\sigma_8=0.9$ (filled circles), LCDM with
$\sigma_8=0.7$ (open triangles), LCDM with $\sigma_8=1.0$ (crosses),
SCDM (filled squares), and OCDM (open circles).  Averaging is
performed over various configurations of triangles with all of scales
of three wavevectors ($k_1$, $k_2$, and $|k_1+k_2|$) less than
$0.03[2\pi/(h^{-1}$ Mpc)] and $\varphi>90^{\circ}$. The error-bars
represent the sample variance of the mock predictions.}
\label{fig:phase_bias_mean}
\end{figure}

Figure \ref{fig:phase_bias_mean} plots $b_2/b_1$ calculated from the
difference between $p^{(3)}_{\rm g}$ and $p^{(3)}_{\rm m}$ for each
cosmology using equation (\ref{eq:p3bias}).  The plotted values of
$b_2/b_1$ are averaged over scales larger than $30h^{-1}$Mpc
($k<0.03[2\pi/(h^{-1}$Mpc)]).  In constraining $b_2/b_1$, we do not 
use the phase sum for the configuration of triangle wavevectors 
with the opening angle, $\varphi<90^{\circ}$,
because we find that they strongly suffer from the observational 
systematic due to the survey geometry beyond the sample variance. We
note that the ratio of the biasing parameters in Figure
\ref{fig:phase_bias_mean} is calculated from the density field in
redshift space and thus their values are quantitatively different in
real space. 

%%%%%%%%%%%%%%%%%%%%%%%%%%%%%%%%%%%%%%%%%%%%%%%%%%%%%%%
%%%%%%%%%%%%%%%%%%%%%%%%%%%%%%%%%%%%%%%%%%%%%%%%%%%%%%%
%%%%%%%%%%%%%%%%%%%%%%%%%%%%%%%%%%%%%%%%%%%%%%%%%%%%%%%
\begin{figure}[tph]
\centering \FigureFile(80mm,80mm){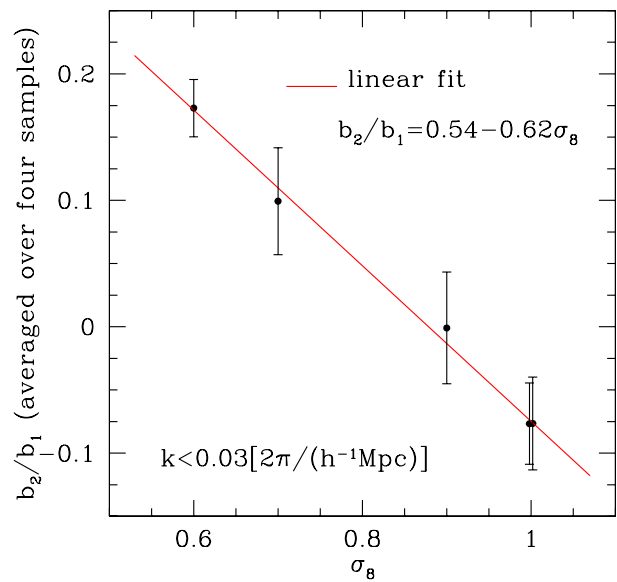}
\caption{Relation between $\sigma_8$ of assumed cosmologies for mock
samples and the averaged $b_2/b_1$ over all four volume-limited
samples under the corresponding cosmologies, which are plotted in
Figure \ref{fig:phase_bias_mean}.  The plotted line represents the
best fit of the above relation with a linear function,
$b_2/b_1=A+B\sigma_8$, where $A=0.54(\pm 0.06)$ and $B=-0.62(\pm
0.08)$.  The error-bars represent the sample variance of the mock
predictions.}
\label{fig:sigma8vsbias}
\end{figure}
%%%%%%%%%%%%%%%%%%%%%%%%%%%%%%%%%%%%%%%%%%%%%%%%%%%%%%%

The results in Figure \ref{fig:phase_bias_mean} suggest that the
nonlinearity of the galaxy biasing, $b_2/b_1$, is generally rather
small and simply dependent on $\sigma_8$, regardless of the cosmology
and galaxy luminosity. We thereby plot the relation between $\sigma_8$
and $b_2/b_1$ averaged over all of four volume-limited samples in
Figure \ref{fig:sigma8vsbias}. Indeed, we find that $\sigma_8$ mainly
determines the resulting $b_2/b_1$ regardless of cosmology at scales
$>30h^{-1}$Mpc, and that their relation is well fitted by the following
linear relation:
%%%%%%%%%%%%%%%%%%%%%%%%%%%%%%%%%%%%%%%%%%%%%%%%%%%%%%%
\begin{equation}
b_2/b_1=(0.54\pm 0.06)-(0.62\pm 0.08)\sigma_8.
\end{equation}
%%%%%%%%%%%%%%%%%%%%%%%%%%%%%%%%%%%%%%%%%%%%%%%%%%%%%%%
The linear relation is valid if $|b_2/b_1|\ll 1$ and the power
spectrum of the mass fluctuation is mainly determined by the square of
$\sigma_8$ at $>30h^{-1}$Mpc; both $p_{\rm m}^{(3)}$ and $f$ (equation
\ref{eq:p3bias}) are roughly given by $\sqrt{P/V_{\rm samp}}$, which
is proportional to $\sigma_8$ in a nearly linear regime. Thus,
$b_2/b_1$ is linearly related to $\sigma_8$ as long as $|b_2/b_1|$ is
much smaller than unity.

When we adopt a value of $\sigma_8\sim 0.9$, measured from the
combined analysis of recent WMAP results and the power spectrum of
SDSS galaxies \citep{Spergel2003, Tegmark2004b},
$b_2/b_1=-0.02\pm 0.1$.  Therefore, we conclude that the galaxy biasing
is well approximated by a linear relation. Complementary measurements
of $\sigma_8$ from other observations which are independent of the
galaxy biasing, including weak gravitational lensing and galaxy
clusters, will put additional constraints on the nonlinearity of the
galaxy biasing.

\section{Summary and Conclusions}

We have performed the first measurement of the distribution function
of the phase sum of galaxies.  We applied the phase statistics to
volume-limited samples of the latest SDSS galaxy catalogs. We found
that the distribution functions of the phase sum for all of the SDSS
galaxy samples are in good agreement with the lowest-order
perturbation formula. Since the latter is characterized by $p^{(3)}$,
we explored various properties of $p^{(3)}$.  For a quantitative
comparison with observations, we constructed realistic mock catalogs
from $N$-body simulations.  We found that the observational phase
correlations agree best with the LCDM model with $\sigma_8=0.9$ if
galaxy biasing is linear; the SCDM model predicts weaker correlations
of phases than observations overall, mainly due to the small
$\sigma_8$.  Actually, we found a systematic increase of $p^{(3)}$ as
$\sigma_8$ increases with the same parameters of the LCDM model. 

Assuming that galaxy biasing is expressed by a quadratic deterministic
relation in a weakly nonlinear regime ($k<0.03[2\pi/(h^{-1}{\rm
Mpc})]$), we placed constraints on a linear combination of the
nonlinearity in the galaxy biasing $b_2/b_1$ and $\sigma_8$ as
$b_2/b_1=0.54(\pm 0.06)-0.62(\pm 0.08)\sigma_8$. Using a recent
measurements of $\sigma_8\sim 0.9$ from a combined analysis of WMAP
and SDSS, the galaxy biasing was found to be well approximated by a
linear relation.

In the present analysis, we have focused on the behavior of $p^{(3)}$,
which carries phase information contained in the bispectrum
\citep{Scocci2001,Scocci2000}. In this sense, our analysis is
equivalent to an analysis of the bispectrum, which is also a first
attempt using SDSS data. Higher order information beyond
bispectrum can be obtained by measuring the distribution function of
the phase sum over more than four numbers of wavevectors which form
closed polygons. Another approach is to decrease $V_{\rm samp}$,
because the perturbative parameter, $p^{(3)}$, is inversely
proportional to $\sqrt{V_{\rm samp}}$. However, extracting
higher order information is limited mainly by the shot noise, as
discussed in section \ref{sec:phasesum}. Therefore, deeper and denser
surveys than SDSS are required to carry out this attempt in reality.

\bigskip

%%%%%%%%%%%%%%%%%%%%%%%%%%%%%%%%%%%%%%%%%%%%%%%%%%%%%%%%%%%%%%%%%%%%%%

We deeply appreciate I.~Kayo and Y.~P.~Jing who kindly provided a large
set of $N$-body simulation data. C.~H. acknowledges support from a JSPS
(Japan Society for the Promotion of Science)
fellowship. T.~M. acknowledges the support from the Ministry of
Education, Culture, Sports, Science, and Technology, Grant-in-Aid for
Encouragement of Young Scientists (No. 15740151).  The research of
Y.S. was supported in part by Grants-in-Aid for Scientific Research from
the Japan Society for Promotion of Science (Nos.14102004 and 16340053).
Numerical computations were carried out at ADAC (the Astronomical Data
Analysis Center) of the National Astronomical Observatory, Japan
(project ID: yys08), and also at computer facilities at the University
of Tokyo supported by the Special Coordination Fund for Promoting
Science and Technology, Ministry of Education, Culture, Sport, Science
and Technology.

Funding for the Sloan Digital Sky Survey (SDSS) has been provided by
the Alfred P. Sloan Foundation, the Participating Institutions, the
National Aeronautics and Space Administration, the National Science
Foundation, the U.S. Department of Energy, the Japanese
Monbukagakusho, the Max Planck Society, and the HEFCE. The SDSS Web
site is http://www.sdss.org/.

The SDSS is a joint project of The University of Chicago, Fermilab, the
Institute for Advanced Study, the Japan Participation Group, The Johns
Hopkins University, the Korean Scientist Group, Los Alamos National
Laboratory, the Max-Planck-Institute for Astronomy (MPIA), the
Max-Planck-Institute for Astrophysics (MPA), New Mexico State
University, University of Pittsburgh, University of Portsmouth,
Princeton University, the United States Naval Observatory, and the
University of Washington.


\bigskip

\begin{thebibliography}{}

\bibitem[Abazajian et al.(2003)]{Abazajian2003} 
Abazajian, K., et al. 2003, \aj, 126, 2081 (Data Release One)

\bibitem[Abazajian et al.(2004)]{Abazajian2004} 
Abazajian, K., et al. 2004, \aj, 128, 502 (Data Release Two)

\bibitem[Abazajian et al.(2005)]{Abazajian2005} 
Abazajian, K., et al. 2005, \aj, 129, 1755 (Data Release Three)

\bibitem[Bardeen et al.(1986)]{BBKS1986}
Bardeen,~J.~M., Bond,~J.~R., Kaiser,~N., \& Szalay,~A.~S. 1986, 
\apj, 304, 15 

\bibitem[Barrow et al.(1985)]{BBS1985}
Barrow, J. D., Bahavsar, S. P., \& Sonoda, D. H. 1985, \mnras, 216, 17

\bibitem[Bertschinger(1992)]{Bertschinger1992}
Bertschinger,~E., 1992 in Lecture Notes in Physics, 408,
New Insights into the Universe, ed. V.~Martinez,
M.~Portilla, \& D.~Saez (Berlin:Springer-Verlag), 65

\bibitem[Blanton et al.(2003a)]{Blanton2003a}
Blanton,~M.~R., Lin,~H., Lupton,~R.~H., Maley,~F.~M., Young,~N.,
Zehavi,~I., \& Loveday,~J. 2003, \aj, 125, 2276

\bibitem[Blanton et al.(2003b)]{Blanton2003b}
Blanton,~M.~R., et al. 2003, \aj, 125, 234

\bibitem[Blanton et al.(2005)]{Blanton2005}
Blanton,~M.~R., et al. 2005, \aj, 129, 2562

\bibitem[Chiang(2001)]{Chiang2001}
Chiang,~L-Y. 2001, \mnras, 325, 405

\bibitem[Chiang et al.(2002)]{CCN2002}
Chiang,~L-Y., Coles,~P., \& Naselsky,~P. 2002, \mnras, 337, 488

\bibitem[Coles, Chiang(2000)]{CC2000}
Coles,~P., \& Chiang,~L-Y. 2000, \nat, 406, 376

\bibitem[Dressler(1980)]{Dressler1980}
Dressler, A. 1980, \apj, 236, 351

\bibitem[Eisenstein et al.(2001)]{Eisen2001}
Eisenstein, D. J., et al. 2001, \aj, 122, 2267

\bibitem[Fukugita et al.(1996)]{Fukugita1996}
Fukugita,~M., Ichikawa,~T., Gunn,~J.~E., Doi,~M., Shimasaku,~K., \&
Schneider,~D.~P. 1996, \aj, 111, 1748 

\bibitem[Gott et al.(1986)]{GMD1986}
Gott,~J.~R.,~III, Melott,~A.~L., \& Dickinson,~M.\ 1986, \apj, 306, 341

\bibitem[Gunn et al.(1998)]{Gunn1998}
Gunn,~J.~E., et al. 1998, \aj, 116, 3040

\bibitem[Hamilton, Tegmark(2004)]{Hamilton2004}
Hamilton,~A.~J.~S., \& Tegmark,~M. 2004, \mnras, 349, 115

\bibitem[Hikage et al.(2002)]{Hikage2002} 
Hikage,~C., 2002, \pasj, 54, 707

\bibitem[Hikage et al.(2003)]{Hikage2003} 
Hikage,~C., 2003, \pasj, 55, 911

\bibitem[Hikage et al.(2004)]{Hikage2004} 
Hikage,~C., Matsubara,~T., \& Suto,~Y. 2004, \apj, 600, 553

\bibitem[Hogg et al.(2001)]{Hogg2001}
Hogg,~D.~W., Finkbeiner,~D.~P., Schlegel,~D.~J., \& Gunn,~J.~E. 2001, 
\aj, 122, 2129

\bibitem[Hoyle et al.(2002)]{Hoyle2002}
Hoyle,~F., et al. 2002, \apj, 580, 663

\bibitem[Ivezi$\acute{\rm c}$ et al. (2004)]{Ivezic2004}
Ivezi$\acute{\rm c}$,~$\breve{\rm Z}$., et al. 2004, Astron. Nachr.,
325, 583

\bibitem[Jain, Bertschinger(1998)]{Jain1998}
Jain,~B., \& Bertschinger,~E. 1998, \apj, 509, 517

\bibitem[Jing and Suto(1998)]{JS1998}
Jing,~Y.~P., \& Suto,~Y. 1998, \apj, 494, L5

\bibitem[Kayo et al.(2004)]{Kayo2004}
Kayo,~I., et al. 2004, \pasj, 56, 415

\bibitem[Komatsu et al.(2003)]{Komatsu2003}
Komatsu,~E., et al. 2003, \apjs, 148, 119

\bibitem[Lupton et al.(2001)]{Lupton2001} 
Lupton,~R., Gunn,~J.~E., Ivezi$\acute{\rm c}$,~$\breve{\rm Z}$.,
Knapp,~G.~R., Kent,~S., \& Yasuda,~N. 2001, in ASP Conf. Ser. 238,
Astronomical Data Analysis Software and Systems X,
ed. F.~R.~Harnden,~Jr., F.~A.~Primini, and H.~E.~Payne 
(San Francisco: Astr. Soc. Pac.) 269

\bibitem[Matsubara(2003)]{Matsu2003}
Matsubara,~T. 2003, \apj, 591, L79

\bibitem[Mecke et al.(1994)]{MBW1994}
Mecke,~K.~R., Buchert,~T., \& Wagner,~H. \ 1994, A\&A, 288, 697

\bibitem[Park et al.(2005)]{Park2005}
Park,~C., et al. 2005, \apj, in press (astro-ph/0507059)

\bibitem[Peebles(1980)]{Peebles1980}
Peebles,~P.~J.~E. 1980, The Large-Scale Structure of the Universe
(Princeton: Princeton University Press)

\bibitem[Pier et al.(2003)]{Pier2003} 
Pier,~J., Munn,~J.~A., Hindsley,~R.~B., Hennessy,~G.~S., Kent,~S.~M.,
Lupton,~R.~H., Ivezi$\acute{\rm c}$,~$\breve{\rm Z}$., for the SDSS
collaboration, 2003, \aj, 125, 1559

\bibitem[Ryden, Gramann(1991)]{Ryden1991}
Ryden,~B.~S., \& Gramann,~M. 1991, \apj, 383, L33

\bibitem[Scherrer et al.(1991)]{SMS1991}
Scherrer,~R.~J., Melott,~A.~L., \& Shandarin,~S.~F. 1991, \apj, 377, 29

\bibitem[Schlegel, Finkbbeiner, and Davis(1998)]{SFD98}
Schlegel,~D.~J., Finkbeiner,~D.~P., \& Davis,~M. 1998, \apj, 500, 525

\bibitem[Scoccimarro(2000)]{Scocci2000}
Scoccimarro,~R. 2000, \apj, 544, 597

\bibitem[Scoccimarro et al.(2001)]{Scocci2001}
Scoccimarro,~R., Feldman,~H.~A., Fry,~J.~N., \& Frieman,~J.~A. 2001,
\apj, 546, 652

\bibitem[Smith et al.(2002)]{Smith2002}
Smith,~J.~A., et al. 2002, \aj, 123, 2121

\bibitem[Soda, Suto(1992)]{Soda1992}
Soda,~J., \& Suto,~Y. 1992, \apj, 396, 379

\bibitem[Spergel et al.(2003)]{Spergel2003}
Spergel,~D.~N., et al. 2003, \apjs, 148, 175 

\bibitem[Stoughton et al.(2002)]{Stoughton2002}
Stoughton,~C., et al. 2002, \aj, 123, 485 (Early Data Release)

\bibitem[Strauss et al.(2002)]{Strauss2002} 
Strauss,~M.~A., et al. 2002, \aj, 124, 1810 

\bibitem[Suginohara, Suto(1991)]{Suginohara1991}
Suginohara,~T., \& Suto,~Y. 1991, \apj, 371, 470

\bibitem[Tegmark et al.(2004a)]{Tegmark2004a}
Tegmark,~M., et al. 2004a, \apj, 606, 702

\bibitem[Tegmark et al.(2004b)]{Tegmark2004b}
Tegmark,~M., et al. 2004b, \prd, 69, 103501

\bibitem[Totsuji and Kihara(1969)]{TK1969}
Totsuji,~H., \& Kihara,~T. 1969, \pasj, 21, 221

\bibitem[Verde et al.(2002)]{Verde2002}
Verde,~L., et al. 2002, \mnras, 335, 432

\bibitem[Watts, Coles(2003)]{WC2003}
Watts,~P., \& Coles,~P. 2003, \mnras, 338, 806

\bibitem[White(1979)]{White1979}
White,~S.~D.~M. 1979, \mnras, 186, 145

\bibitem[Yahata et al.(2005)]{Yahata2005}
Yahata,~K., et al. 2005, \pasj, 57, 529

\bibitem[York et al.(2000)]{York2000} 
York,~D.~G., et al. 2000, \aj, 120, 1579. 

\bibitem[Zehavi et al.(2002)]{Zehavi2002}
Zehavi,~I., et al. 2002, \apj, 571, 172

\bibitem[Zehavi et al.(2005)]{Zehavi2005}
Zehavi,~I., et al. 2005, \apj, in press (astro-ph/0408569)

\end{thebibliography}
\end{document}